\documentclass[aps,prc,superscriptaddress,floatfix,nofootinbib,twocolumn]{revtex4-2}

\usepackage{amsmath,amssymb,amsfonts,physics,graphicx,float}
\graphicspath{{pic/}}
\usepackage[setpagesize=false]{hyperref}
\allowdisplaybreaks[4]

\usepackage[labelformat=simple]{subcaption} 

\usepackage{ulem}
\usepackage{color}
\usepackage{slashed}
\definecolor{ar}{rgb}{1.0, 0.01, 0.24}
\definecolor{al}{rgb}{0.82, 0.1, 0.26}
\definecolor{ev}{rgb}{0.56, 0.0, 1.0}

\newcommand{\lag}{\mathcal{L}}

\newcommand{\UA}[1]{\mathrm{U}(#1)_\mathrm{A}}

\newcommand{\rl}{\mathrm{L}}
\newcommand{\rr}{\mathrm{R}}

\newcommand{\GO}{Gell-Mann--Okubo mass relation}

\begin{document}

\title{
Linear realization of SU(3) parity doublet model for octet baryons with bad diquark
}

\author{Bikai Gao}
\email{bikai@rcnp.osaka-u.ac.jp}
\affiliation{Research Center for Nuclear Physics, Osaka University, Ibaraki, Osaka 567-0047, Japan}

\author{Atsushi Hosaka}
\email{hosaka@rcnp.osaka-u.ac.jp}
\affiliation{Research Center for Nuclear Physics, Osaka University, Ibaraki, Osaka 567-0047, Japan}
\affiliation{Advanced Science Research Center, Japan Atomic Energy Agency, Tokai 319-1195, Japan}

\date{\today}

\begin{abstract}
We construct a  linear $SU(3)_L \times SU(3)_R$ parity doublet model for octet baryons. Our model employs the $(3,\bar{3}) + (\bar{3},3)$ and $(3,6) + (6,3)$ chiral representations while excluding the $(8,1) + (1,8)$ representation. Through systematic analysis, we demonstrate that the $(3,6) + (6,3)$ representation containing symmetric ``bad'' diquarks, despite being energetically disfavored, is essential for reproducing the correct baryon mass hierarchy—particularly the $\Sigma$-$\Xi$ mass ordering. The model incorporates both spontaneous and explicit chiral symmetry breaking, with the latter implemented through bare quark mass terms that properly account for $SU(3)$ flavor breaking effects. Our numerical analysis successfully reproduces the ground-state octet baryon masses and predicts the spectrum of excited states up to 2.5 GeV. For the experimentally challenging $\Xi$ sector, we identify $\Xi(1950)$ as the first positive-parity excitation. The analysis reveals that ground states are dominated by the $(3,\bar{3}) + (\bar{3},3)$ representation, consistent with the preference for ``good'' diquark configurations, while the $(3,6) + (6,3)$ contribution remains crucial for the mass spectrum. 
\end{abstract}

\maketitle


\section{Introduction}

The study of baryon masses and their chiral structure remains one of the fundamental challenges in hadron physics. While the traditional view attributes baryon masses primarily to spontaneous chiral symmetry breaking through quark condensates, theoretical frameworks based on parity doublet structure that incorporate both chiral-variant and chiral-invariant mass terms have been developed to provide a more complete description of baryon properties~\cite{Detar:1988kn, Jido:1998av, Jido:2001nt, Nagata:2008xf, Sasaki:2010bp, Gallas:2013ipa}. This theoretical framework has received strong support from recent lattice QCD studies, which have revealed the existence of chiral-invariant mass contributions that persist even at the chiral restoration point~\cite{Aarts:2015mma, Aarts:2017rrl, Aarts:2018glk}. Furthermore, lattice QCD calculations of axial charges for negative-parity nucleon resonances have shown that $N^*(1535)$ has a remarkably small axial charge $g_A \sim \mathcal{O}(0.1)$, while $N^*(1650)$ exhibits $g_A \sim 0.5$, providing important insights into the chiral structure of excited baryons~\cite{Takahashi:2008fy}.

In the context of three-flavor QCD, two main approaches have emerged for implementing chiral symmetry in effective models. The nonlinear realization, where pions appear as Nambu-Goldstone bosons with derivative couplings, has been extensively studied in the framework of chiral perturbation theory and various effective models~\cite{Weinberg:1968de,Bando:1987br,Buscher:1987qj,Gasser:1987zq,Coleman:1969sm,Callan:1969sn}. Several authors have developed $SU(3)$ chiral models with parity doublet structure using this nonlinear realization~\cite{Papazoglou:1998vr,Mishra:2003tr,Steinheimer:2011ea,Dexheimer:2012eu,Fraga:2023wtd}. While these models have achieved success in describing nuclear matter and the phase structure at finite density, they face inherent limitations when extended to energy scales approaching the chiral restoration region, where the derivative expansion becomes questionable.

In contrast, the linear realization of chiral symmetry, where chiral partners are explicitly included as dynamical degrees of freedom, offers a more suitable framework for studying physics near and beyond the chiral restoration scale~\cite{Hatsuda:1988mv, Manohar:1983md,Zschiesche:2006zj,Dexheimer:2007tn,Chen:2009sf,Chen:2010ba,Sasaki:2010bp,Chen:2011rh,Sasaki:2011ff,Motohiro:2015taa,Benic:2015pia,Nishihara:2015fka,Suenaga:2017wbb,Takeda:2017mrm,Marczenko:2017huu,Mukherjee:2017jzi,Marczenko:2018jui,Yamazaki:2019tuo,Harada:2019oaq,Harada:2020etl,Marczenko:2020yok,Marczenko:2021icv}. However, despite its theoretical advantages, relatively little work has been done on $SU(3)_L \times SU(3)_R$ linear sigma models with parity doublet structure for baryons. The increased complexity arising from the numerous possible chiral representations and the need to properly implement both spontaneous and explicit symmetry breaking has made systematic studies challenging~\cite{Nishihara:2015fka,Minamikawa:2023ypn,Gao:2024mew, Christos:1982kc}.

Constructing a systematic and predictive linear model for the baryon octet presents unique challenges. The $SU(3)_L \times SU(3)_R$ chiral symmetry allows for numerous possible representations for baryons, making the selection of relevant degrees of freedom a crucial theoretical question. From the perspective of quark dynamics, baryons can be viewed as composed of a quark and a diquark, where the diquark structure plays an essential role in determining the baryon properties~\cite{Jaffe:1976ig,Jaffe:1976ih,Rapp:1997zu,Jaffe:2003sg,Eichmann:2016yit,Chen:2022asf}.

In our previous work~\cite{Minamikawa:2023ypn}, we attempted to construct a minimal linear model using only the $(3_L,\bar{3}_R) + (\bar{3}_L,3_R)$ and $(8_L,1_R) + (1_L,8_R)$ representations, which contain ``good'' (antisymmetric) diquarks. However, even after including second-order interactions in the scalar field $M$, this approach failed to reproduce the correct mass hierarchy of the baryon octet, particularly the $\Sigma$-$\Xi$ mass ordering. This limitation suggested that representations containing ``bad'' (symmetric) diquarks might be necessary for a complete description.

To address this issue, we subsequently developed a comprehensive model in Ref.~\cite{Gao:2024mew} that included all relevant representations: $(3_L,\bar{3}_R) + (\bar{3}_L,3_R)$, $(8_L,1_R) + (1_L,8_R)$, and $(3_L,6_R) + (6_L,3_R)$. While this approach successfully reproduced the mass spectrum, it suffered from significant drawbacks. The model required ten coupling parameters, which introduced substantial ambiguities in parameter determination due to the presence of local minima and the lack of experimental data for higher excited states. Moreover, the analysis was performed only with spontaneous chiral symmetry breaking, neglecting explicit breaking effects from quark masses, which are crucial for understanding $SU(3)$ flavor breaking patterns.

In this work, we pursue a different strategy by constructing a simplified linear $SU(3)_L \times SU(3)_R$ model that balances predictive power with theoretical clarity. We focus on the $(3_L,\bar{3}_R) + (\bar{3}_L,3_R)$ and $(3_L,6_R) + (6_L,3_R)$ representations, deliberately excluding the $(8_L,1_R) + (1_L,8_R)$ representation. This choice is motivated by several considerations. First, the $(3,\bar{3})$ representation naturally contains the flavor-antisymmetric diquark structure favored by one-gluon exchange interactions. Second, as demonstrated in our previous work, the $(3,6)$ representation with symmetric diquarks is essential for reproducing the correct mass ordering, particularly for excited states. By excluding the $(8,1)$ representation, we significantly reduce the number of parameters with revealing the essential feature for the  mass spectrum of SU(3) baryons including excited states. Crucially, we incorporate explicit chiral symmetry breaking through quark mass terms, which was absent in our previous comprehensive analysis. This allows us to properly account for $SU(3)$ flavor breaking effects and study the interplay between spontaneous and explicit symmetry breaking in the linear realization framework. The explicit breaking terms are essential for understanding the mass differences between strange and non-strange baryons and for making meaningful comparisons with experimental data.

Our simplified linear model provides a tractable framework for studying baryon properties across different density regimes, from vacuum to dense matter relevant for neutron stars, where the linear realization becomes increasingly important as the system approaches chiral restoration. The reduced parameter space also allows for more systematic exploration of the model's predictions and clearer physical interpretation of the results.

The remainder of this paper is organized as follows. In Sec.~\ref{sec-representation}, we review the chiral representation theory for hadrons and establish the connection between quark-diquark structure and specific $SU(3)_L \times SU(3)_R$ representations. Section~\ref{sec_3} systematically examines leading-order Yukawa interactions for different chiral representations, demonstrating through explicit calculation why the $(3,\bar{3}) + (\bar{3},3)$ representation alone produces unphysical degeneracies, why adding $(8,1) + (1,8)$ yields incorrect mass orderings~\cite{Minamikawa:2023ypn}, and why the $(3,6) + (6,3)$ representation is essential despite containing energetically disfavored diquarks. In Sec.~\ref{sec_4}, we construct the complete parity doublet model incorporating both $(3,\bar{3}) + (\bar{3},3)$ and $(3,6) + (6,3)$ representations with their mirror assignments, including both spontaneous and explicit chiral symmetry breaking terms. Section~\ref{sec_5} presents our numerical analysis, where we fit the model parameters to experimental baryon masses and examine the resulting mass spectrum, baryon composition, and model predictions for excited states. We conclude with a summary and discussion in Sec.~\ref{sec-summary} and appendices provide additional  details.


\section{Chiral representations for baryons}\label{sec-representation}
Quarks transform under $SU(3)_L \times SU(3)_R \times U(1)_A$ symmetry as
\begin{equation}
\begin{aligned}
&q_{L} \rightarrow  e^{-i \theta_{A}} g_{L}q_{L}, \\
&q_{R} \rightarrow  e^{+i \theta_{A}} g_{R}q_{R},
\end{aligned}
\end{equation}
with $g_{L,R} \in$ SU(3)$_{L,R}$ and $\theta_{A}$ being the $U(1)_A$ transformation parameters. 
Accordingly, we assign the U(1)$_{A}$ charge of the left and right handed quarks as $-1$ and $+1$, respectively. The chiral representation of the left and right handed quark is then given by
\begin{equation}
q_{L} :( 3, 1 )_{-1}, \quad q_R :(1 , 3 )_{+1}
\end{equation}
where these $3$ and $1$ in the bracket express the triplet and singlet for SU(3)$_{L}$ symmetry and SU(3)$_{R}$ symmetry, respectively. 
The index indicates the axial charge of the fields. 

Since baryons consist of three valence quarks, 
the baryon fields are formed by the tensor products of three quark fields. 
We define the left-handed baryon field 
as a product of a spectator left-handed quark 
and left- or right-handed diquark, 
while the right-handed baryon has a right-handed spectator quark. 
Taking irreducible decomposition, the left-handed baryon 
can be expressed as the following representations 
\begin{align}
q_\rl&\otimes(q_\rl\otimes q_\rl+q_\rr\otimes q_\rr)\notag\\
\sim&(1,1)_{-3}+(8,1)_{-3}+(8,1)_{-3}\notag\\
&+(10,1)_{-3} +(3,\bar3)_{+1}+(3,6)_{+1}\,. 
\end{align}
After the chiral symmetry is spontaneously broken down to flavor $SU(3)_F$ symmetry, octet baryons appear from the representations of $(3,\bar3)$, $(8,1)$, and $(3,6)$. The representations $(3,\bar3)$  contain flavor-antisymmetric diquarks $\sim\bar3$ which is called ``good'' diquark,  
while $(3,6)$ contains flavor-symmetric diquark $\sim6$ called ``bad'' diquark. Also, for simplicity, we only focus on baryons having spin 1/2. 

The diquark picture provides essential insights into baryon structure and dynamics. While baryons fundamentally consist of three valence quarks, the complexity of the three-body problem in QCD makes direct calculations extremely challenging. The emergence of diquark correlations offers a crucial simplification: the intricate three-quark system reduces to a more tractable quark-diquark configuration, where two quarks form a correlated pair that together with the third spectator quark. This reduction reflects important physical dynamics rather than mere mathematical convenience. The one-gluon exchange interaction creates an attractive force in the color-antitriplet channel, naturally favoring diquark formation. Among these correlations, ``good" diquarks in the flavor-antisymmetric $\bar{3}$ representation are particularly favored—they benefit from both attractive color forces and  spin-flavor correlations, resulting in compact, tightly bound configurations. In contrast, ``bad" diquarks with symmetric flavor structure experience weaker binding despite the same color attraction.

These different binding strengths directly manifest in the baryon mass spectrum. Baryons dominated by good diquark correlations systematically exhibit lower masses than those containing bad diquarks, explaining key features of the mass hierarchy within baryon multiplets. The diquark framework thus bridges microscopic quark-gluon dynamics and macroscopic baryon properties, providing both computational tractability and physical transparency for understanding experimental observations.
\section{Yukawa interactions at the leading-order}\label{sec_3}

In this section, we construct Yukawa-type meson-baryon Lagrangians
using suitable chiral representations of baryons. We begin with the simplest possible model containing only the $(3,\bar{3}) + (\bar{3},3)$ representation, which includes the favored ``good'' diquark structure. When this minimal approach fails to generate the correct mass spectrum---specifically producing an unphysical degeneracy between nucleon and $\Xi$ baryons---we extend the model by adding the $(8,1) + (1,8)$ representation. However, this extension, while resolving the degeneracy issue, introduces a new problem: it predicts an incorrect mass ordering with the nucleon heavier than the $\Sigma$ baryon. Through this systematic analysis, we introduce the $(3,6) + (6,3)$ representation containing ``bad'' diquarks, despite being energetically less favorable, is essential for achieving the correct octet baryon mass hierarchy. 

\subsection{($3, \bar{3}$) + ($\bar{3}, 3$) representation} \label{sec_3_3}
\begin{figure}\centering
\includegraphics[width=1\hsize]{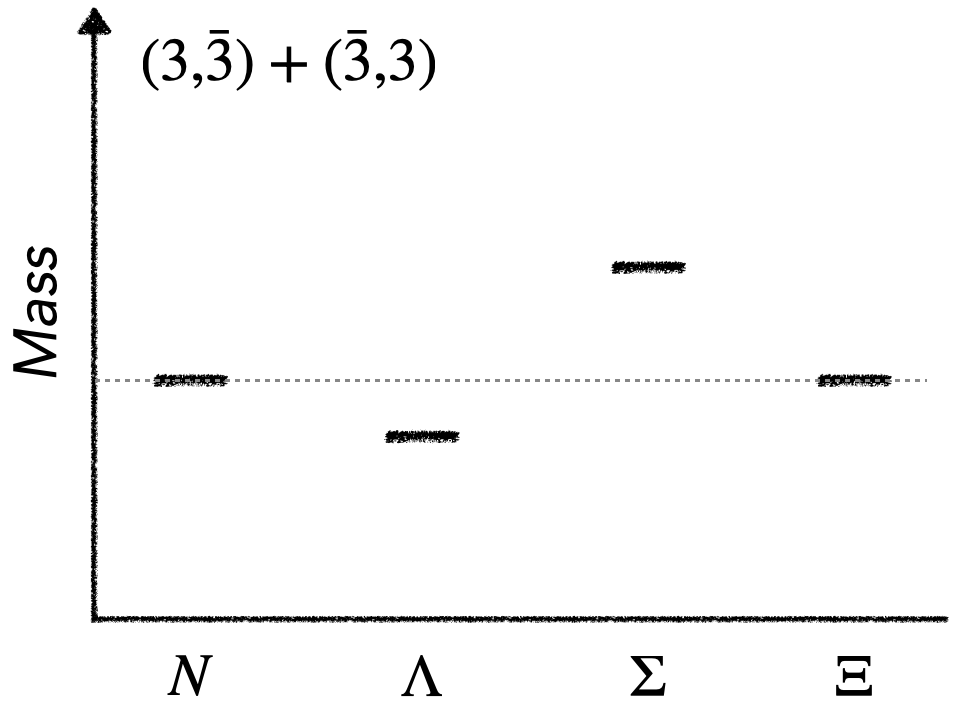}
\caption{Phenomenological mass spectrum for the case with only $(3, \bar{3}) + (\bar{3}, 3)$ representation. In this scenario, there is a single physical state with one parameter $g$, and the mass ordering is $m_{\Sigma} > m_{N} = m_{\Xi} >  m_{\Lambda}$.  }
\label{fig_m_3_3bar}
\end{figure}

Let us first consider a simple model including the (3, $\bar{3}$) + ($\bar{3}$, 3) representations for octet baryons at the leading order. We introduce the baryon field $\psi$ correspond to the representations as
\begin{align}
\psi_{L} \sim (3, \bar{3})_{+1}, \quad \psi_{R} \sim (\bar{3}, 3)_{-1}.
\end{align}
with chiral transformation properties as
\begin{align}
\psi_{L} \rightarrow g_L \psi_{L} g_{R}^{\dagger}, \quad \psi_{R} \rightarrow g_{R}\psi_{R} g_{L}^{\dagger}.
\end{align}
The explicit component of the $\psi$ field is 
\begin{align}
(\psi)^i_j&\sim
\frac1{\sqrt3}\Lambda_0 + (B)^i_j \\
(B)^i_j &=\begin{bmatrix}
\frac1{\sqrt2}\Sigma^0+\frac1{\sqrt6}\Lambda & \Sigma^+ & p \\
\Sigma^- & -\frac1{\sqrt2}\Sigma^0+\frac1{\sqrt6}\Lambda & n \\
\Xi^- & \Xi^0 & -\frac2{\sqrt6}\Lambda \\
\end{bmatrix}\,, \label{eq_octet}
\end{align}
for left-handed and right-handed respectively.
We also introduce a $3 \times 3$ matrix field $M$ for scalar and pseudo-scalar mesons as
\begin{align}
M \sim (3 , \bar{3})_{-2}.
\end{align}
This field is made from $(q_L\bar{q}_R)$ and have the chiral transformation property as
\begin{align}
M \rightarrow g_L M g_{R}^{\dagger}.
\end{align}
By taking the mean-field approximation, the meson field is expressed as
\begin{align}
\langle M \rangle = {\rm diag}(\alpha, \beta, \gamma).
\end{align}
Under the assumption of isospin symmetry, we set $\alpha = \beta = f_\pi$. To incorporate explicit flavor symmetry breaking effects, we choose $\gamma = 2f_K - f_\pi$, where $f_\pi$ and $f_K$ are the pion and kaon decay constants with the explicit values shown in Table.~\ref{tab-condensate-input}. These parameter choices naturally encode the mass difference between strange and non-strange quarks through the meson field expectation values.
\begin{table}
\caption{
Physical inputs of the decay constants for pion and kaon,
and the VEV of the meson field  $\langle M \rangle={\rm diag}\{{\alpha,\beta,\gamma}\}$. 
}
\label{tab-condensate-input}
\centering
\begin{tabular}{c|c}
\hline\hline
$f_\pi$ & 93 MeV \\
$f_K$ & 110 MeV \\
$\alpha=\beta$ & $f_\pi(=93\,\mathrm{MeV})$ \\
$\gamma$ & $2f_K-f_\pi(=127\,\mathrm{MeV})$ \\
\hline\hline
\end{tabular}
\end{table}
With these configurations, the leading-order chiral invariant Yukawa interactions can be constructed as
\begin{align}
\lag_1
=g\big[&\varepsilon^{l_1l_2l_3}\varepsilon_{r_1r_2r_3}
(\bar\psi_L)^{r_1}_{l_1}(M^\dag)^{r_2}_{l_2}(\psi_R)^{r_3}_{l_3}\notag\\
+&\varepsilon_{l_1l_2l_3}\varepsilon^{r_1r_2r_3}
(\bar\psi_R)^{l_1}_{r_1}(M)^{l_2}_{r_2}(\psi_L)^{l_3}_{r_3}\big]\label{eq-case1}\,, 
\end{align}
where $\varepsilon_{ijk}$ is the totally antisymmetric tensor. However, this interaction term, which originates from the chiral structure, suffers from a serious problem. In this simple model, the strange quark condensate $\sigma_s$, corresponding to $\langle \bar{s}s \rangle$, is located in the $(3,3)$ component of the meson field $(M)^3_3$. Consider the second term as an example:
\begin{align}
\varepsilon_{l_1 3 l_3}\varepsilon^{r_1 3 r_3}
(\bar\psi_R)^{r_1}_{l_1}(M^\dagger)^{3}_{3}(\psi_L)^{r_3}_{l_3}\, .
\end{align}
Due to the antisymmetric nature of the tensor, the indices $l_1, l_3$ and $r_1, r_3$ cannot take the value 3, which prevents this term from contributing. Consequently, the strange quark condensate $\sigma_s$ contributes only to the $\Sigma$ and octet $\Lambda$ baryons, while leaving the nucleon and $\Xi$ baryon masses degenerate. 
The mass matrix for each baryon is obtained as
\begin{equation}
\begin{aligned}
\hat{M}_N = g \alpha,& \quad \hat{M}_{\Sigma} = g \gamma,\\
\hat{M}_{\Xi} = g \alpha,& \quad  \hat{M}_{\Lambda} = -\frac{g}{3}(\gamma - 4 \alpha) 
\end{aligned}
\end{equation}
with the corresponding phenomenological mass spectrum is shown in Fig.~\ref{fig_m_3_3bar}. These mass matrix satisfy the Gell-Mann-Okubo relation as will be reviewed in Sec.~\ref{sec_5}.
To resolve this mass degeneracy and recover the correct mass hierarchy, additional chiral representations must be incorporated.

To clarify the origin of the wrong mass ordering, we examine the flavor symmetry structure of the chiral Lagrangian with the $(3, \bar{3}) + (\bar{3}, 3)$ representation. To understand the interaction structure, we decompose the second term in Eq.~(\ref{eq-case1}) into trace terms using the identity between the Levi-Civita tensor and Kronecker delta:
\begin{equation}
\varepsilon_{l_1 l_2 l_3} \varepsilon^{r_1 r_2 r_3}=\operatorname{det}\left[\begin{array}{lll}
\delta_{l_1}^{r_1} & \delta_{l_1}^{r_2} & \delta_{l_1}^{r_3} \\
\delta_{l_2}^{r_1} & \delta_{l_2}^{r_2} & \delta_{l_2}^{r_3} \\
\delta_{l_3}^{r_1} & \delta_{l_3}^{r_2} & \delta_{l_3}^{r_3}
\end{array}\right].
\end{equation}
Applying this identity yields
\begin{align}\label{eq_decom}
&\varepsilon_{l_1l_2l_3}\varepsilon^{r_1r_2r_3}
(\bar\psi_R)^{l_1}_{r_1}(M)^{l_2}_{r_2}(\psi_L)^{l_3}_{r_3} \nonumber\\
&= {\rm tr}( \bar{\psi}_{1R} M \psi_{1L}) +  {\rm tr}(\bar{\psi}_{1R} \psi_{1L} M) +{\rm tr}(\bar{\psi}_{1R}) {\rm tr}(M) {\rm tr}(\psi_{1L}) \nonumber \\
& \quad - {\rm tr}(M){\rm tr}( \bar{\psi}_{1R} \psi_{1L} ) - {\rm tr}(\bar{\psi}_{1R}) {\rm tr}(M \psi_{1L}) \nonumber\\
&\quad - {\rm tr}(\bar{\psi}_{1R}M){\rm tr}(\psi_{1L}).
\end{align}

The contraction through $\delta^{r}_l$ explicitly breaks chiral symmetry, resulting in each term of Eq.~(\ref{eq_decom}) preserving flavor symmetry while violating chiral symmetry. Notably, the first two terms can be combined to form the D-type interaction ${\rm tr}(\bar{\psi} \{ M, \psi \})$, where $\{,\}$ denotes the anticommutator. However, the F-type interaction ${\rm tr}(\bar{\psi} [ M, \psi ])$ is absent when considering only the $(3, \bar{3}) + (\bar{3}, 3)$ representation.

This absence of F-type interactions represents a fundamental limitation of the minimal $(3, \bar{3}) + (\bar{3}, 3)$ model. In the conventional $SU(3)$ quark model, both D-type and F-type couplings are essential for properly describing the baryon mass spectrum and interaction patterns. Therefore, additional chiral representations must be incorporated to achieve a complete description comparable to the $SU(3)$ quark model framework.

\subsection{Adding  (8, 1) + (1, 8) representation}
Next we introduce the representation $\chi$ field with the chiral representation
\begin{align}
\chi_{L} \sim (8, 1)_{-3}, \quad \chi_R \sim (1, 8)_{+3}.
\end{align}
Under chiral transformation, they transform similarly to the $\psi$ field as
\begin{align}
\chi_L \rightarrow g_L \chi_L g_L^{\dagger}, \quad \chi_R \rightarrow g_R \chi_R g_R^{\dagger}.
\end{align}
The components of $\chi$ field is taken as
\begin{equation}
(\chi)^i_j = (B)^i_j.
\end{equation}
We can then construct the first order Yukawa interactions between $\psi$ and $\chi$.
The Lagrangian at the leading order in $M$ is 
\begin{align}
\lag_2
& = \lag_1
+ h \tr\big[
\bar\chi_\rl M\psi_\rr
+\bar\chi_\rr M^\dag\psi_\rl
\big]\,. \label{eq-case2}
\end{align}
We emphasize that, at the first orders in $M$,
there is no Yukawa interactions that couple $\chi_L$ and $\chi_R$ fields.
This is because the $\chi$ contains 
three valence quarks with all left-handed or right-handed
so that Yukawa interactions with $\chi$ should include three quark exchanges
that flip the chirality of all three quarks. 
In other words, since $\UA{1}$-charges for $\chi_\rl$, $\chi_\rr$, $M$ are 
$-3$, $3$, and $-2$ respectively, 
a $\UA{1}$ symmetric term cannot be constructed unless we consider the cubic orders, $M^3$ or $(M^\dag)^3$.

\subsubsection{Mass hierarchies}
Using the VEV of $M$, the lagrangian can be decomposed into the form 
\begin{align}
\begin{aligned}
& 
\lag_2 =
-\begin{pmatrix} \bar\psi_N & \bar\chi_N \\ \end{pmatrix}
\hat{M}_N 
\begin{pmatrix} \psi_N \\ \chi_N \\ \end{pmatrix}
%
-\begin{pmatrix} \bar\psi_\Sigma & \bar\chi_\Sigma \\ \end{pmatrix}
\hat{M}_\Sigma
\begin{pmatrix} \psi_\Sigma \\ \chi_\Sigma \\ \end{pmatrix}\\
& \quad {} -\begin{pmatrix} \bar\psi_\Xi & \bar\chi_\Xi \\ \end{pmatrix}
\hat{M}_\Xi
\begin{pmatrix} \psi_\Xi \\ \chi_\Xi \\ \end{pmatrix}
+(\text{terms for }\Lambda\text{ baryons})
\ ,
\end{aligned}
\end{align}
where $\hat{M}_N$, $\hat{M}_\Sigma$, $\hat{M}_\Xi$ and $\hat{M}_\Lambda$ are $2\times2$ mass matrices given by 
\begin{align}
\begin{aligned}
& \hat{M}_N = 
\begin{pmatrix}
g\alpha & -h\alpha \\
-h\alpha & 0 \\
\end{pmatrix}
\ , 
\\
& \hat{M}_\Sigma = 
\begin{pmatrix}
g\gamma & -h\alpha \\
-h\alpha & 0 \\
\end{pmatrix}
\ , \\
& \hat{M}_\Xi = 
\begin{pmatrix}
g\alpha & -h\gamma \\
-h\gamma & 0 \\
\end{pmatrix}, \\
& \hat{M}_\Lambda = 
\begin{pmatrix}
\frac{g}{3}(4\alpha - \gamma) & -\frac{h}{3}(\alpha + 2\gamma) \\
 -\frac{h}{3}(\alpha + 2\gamma) & 0 \\
\end{pmatrix}
\ .
\end{aligned}
\end{align}

The strange quark contributions to the $\Sigma$ baryons enter the diagonal components of the mass matrix.
The one for $\Xi$ baryons, which enters the off-diagonal components. 
The mass eigenvalues for the ground-state octet members can be written as 
\begin{equation}
\begin{aligned}
m[N]&=m(|g\alpha|,|h\alpha|)\,, \\
m[\Sigma]&=m(|g\gamma|,|h\alpha|)\,, \\
m[\Xi]&=m(|g\alpha|,|h\gamma|)\,, 
\end{aligned}
\end{equation}
where $m(x,y)\equiv\sqrt{(x/2)^2+y^2}-x/2$ is an eigenvalue of the matrix 
$\begin{pmatrix} x & y \\ y & 0 \end{pmatrix}$.

\begin{figure}\centering
\includegraphics[width=1\hsize]{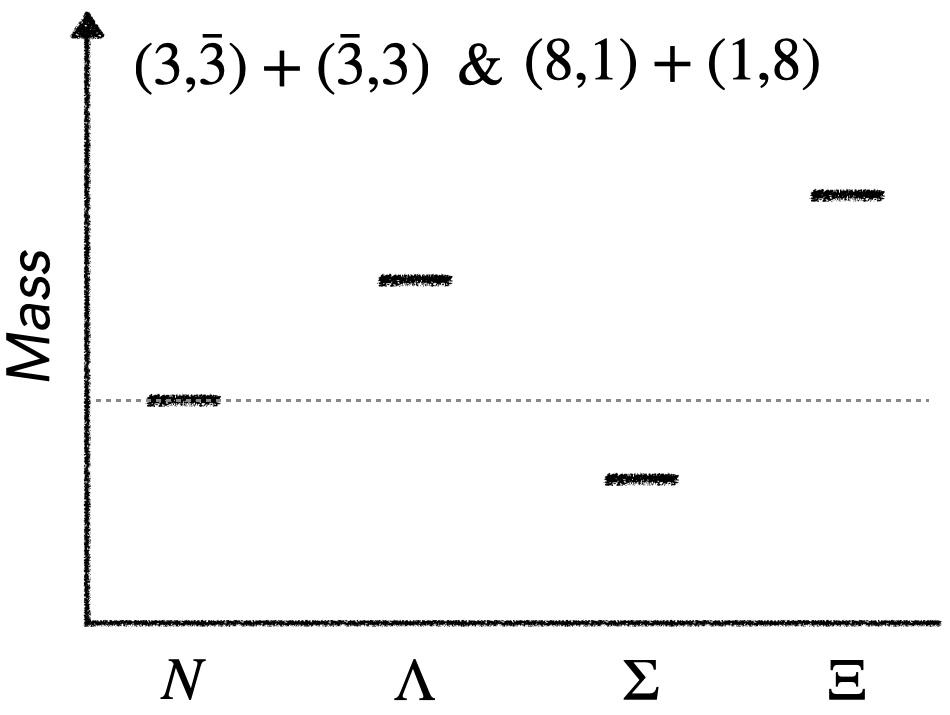}
\caption{Phenomenological mass spectrum for the case with  $(3, \bar{3}) + (\bar{3}, 3)$ and (8 , 1) + (1, 8) representation. In this scenario, with parameters $g$ and $h$ fixed,  the  predicted mass ordering of the ground state baryons is  $m_{\Xi} > m_{\Lambda} > m_{N} >  m_{\Sigma}$.  }
\label{fig_33bar_81}
\end{figure}

Although the addition of the $(8,1) + (1,8)$ representation successfully resolves the mass degeneracy between the nucleon and $\Xi$ baryon, this model still fails to reproduce the correct octet baryon mass spectrum~\cite{Minamikawa:2023ypn}. The underlying issue stems from the parameter hierarchy: since $|g\alpha| < |g\gamma|$ and the mass function satisfies $\partial_x m(x,y) < 0$, the model predicts $m[N] > m[\Sigma]$. The corresponding phenomenological mass spectrum of ground state baryons is shown in Fig.~\ref{fig_33bar_81}  for a certain choice of parameters, $g$ and $h$. 
This prediction contradicts the experimental observation that the nucleon is lighter than the $\Sigma$ baryon. Therefore, despite resolving the degeneracy problem, this extended model cannot accurately reproduce the octet baryon masses.


\subsection{  (3, 6) + (6, 3) representation}

We now discuss the \((3, 6) + (6, 3)\) representation in this subsection. 
The \((3, 6)\) representation contains the so-called ``bad'' diquark, 
which is symmetric in flavor space. Although the color attraction remains, 
this flavor-symmetric configuration is energetically disfavored compared 
to the antisymmetric one. Despite its suppression, the inclusion of this 
configuration is essential for a comprehensive description of baryon 
properties---particularly for reproducing the axial charge. In the \((3, \bar{3})\) representation, the diquark carries zero spin and 
isospin, so only the remaining (spectator) quark contributes to the axial 
current. In a simple quark model calculation, this yields
\begin{equation}
g_A^{(3,\bar{3})} = 1.
\end{equation}
In contrast, within the \((3, 6)\) representation, the symmetric diquark 
allows both quarks to align their spins and isospins coherently, leading to
\begin{equation}
g_A^{(3,6)} = \frac{5}{3},
\end{equation}
which helps recover the experimental value of the axial charge, 
\(g_A^{\mathrm{exp}} \approx 1.27.\)

\begin{figure}\centering
\includegraphics[width=1\hsize]{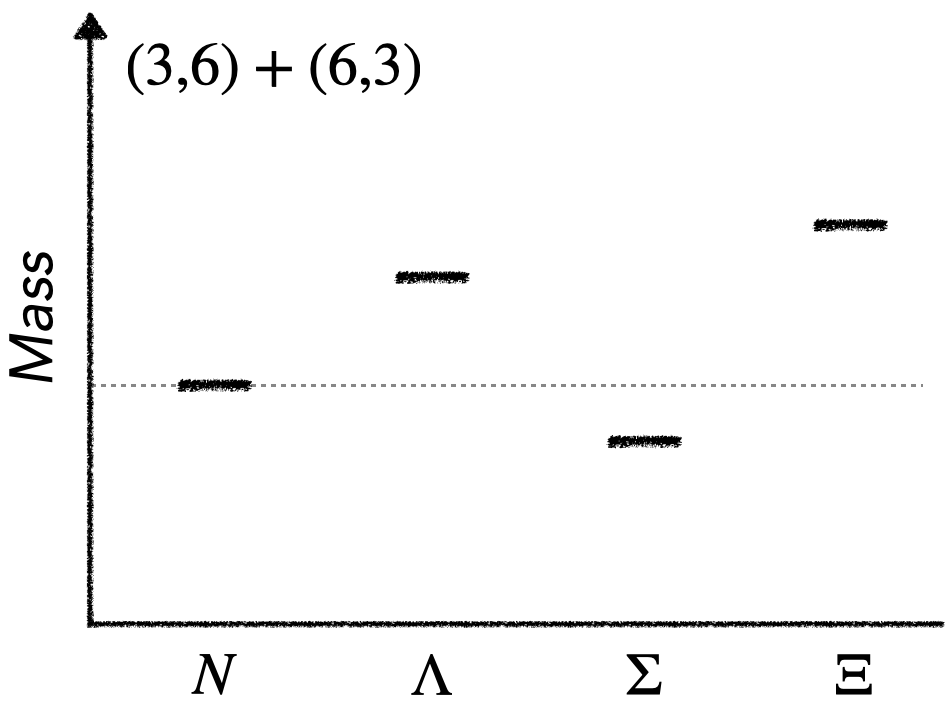}
\caption{Phenomenological mass spectrum for the case with  only (3 , 6) + (6, 3) representation. The  predicted mass ordering of the ground state baryons is  $m_{\Xi} > m_{\Lambda} > m_{N} >  m_{\Sigma}$.  }
\label{fig_33bar_36}
\end{figure}

We introduce the $\eta$ fields correspond to the chiral representation
\begin{align}
\eta_L \sim (3, 6)_{+1}, \quad \eta_R \sim (6, 3)_{-1}.
\end{align}
They transform under the chiral transformation as
\begin{align}
\eta_{L} \rightarrow g_L \eta_L g_R^{\dagger}, \quad \eta_R \rightarrow g_R \eta_R g_L^{\dagger}.
\end{align}
 For clarifying the representations under chiral group, we explicitly write the superscripts of the $\eta$ fields as
\begin{align}
\eta_L^{(a , \alpha \beta)}, \quad \eta_R^{ (ab, \alpha)}
\end{align}
where $a, b = 1, 2, 3$ are for $SU(3)_L$ and $\alpha = 1, 2, 3$ for $SU(3)_R$. The superscripts of $ab$ and $\alpha \beta$ are symmetrized to express 6 representation. The components of the $\eta$ field is 
\begin{align}
\eta_L^{(a , \alpha \beta)} &= \Delta_L^{a \alpha \beta} + \frac{1}{\sqrt{6}}\left( \epsilon^{\alpha ac} \delta_k^\beta + \epsilon^{\beta a c} \delta_k^\alpha \right) (B_L)^k_c, \\
\eta_R^{(ab , \alpha )} &= \Delta_R^{ab \alpha} + \frac{1}{\sqrt{6}}\left( \epsilon^{a\alpha c} \delta_k^b + \epsilon^{b \alpha c} \delta_k^a \right) (B_R)^k_c, 
\end{align}
with $\Delta, B$ the decuplet and the octet baryons. In this work, we simply neglect the decuplet states and focus on the octet baryons. We can then construct the first order Yukawa interactions between $\psi$ and $\eta$ at the leading order of $M$ 
\begin{align}
\mathcal{L}_{{\rm \psi\eta}} &=  y \left[\epsilon_{abc} (\bar{\psi}_{1R})^a_\alpha (M)_\beta^b (\eta_{1L})^{(c, \alpha \beta)} + {\rm h.c.} \right.  \nonumber\\
&\quad \left.  \epsilon_{\alpha \beta \sigma} (\bar{\psi}_{1L})^\alpha_a (M^\dagger)^\beta_b (\eta_{1R})^{(ab, \sigma)}    + {\rm h.c.}  \right]. 
\end{align}
and the self-interaction term 
\begin{align}
\mathcal{L}_{\eta_{\rm self}} = g_\eta \left[ (\bar{\eta}_{1R})_{(ab, \alpha)} (M)^a_\beta (\eta_{1L})^{ (b, \alpha \beta)} + {\rm h. c.} \right] 
\end{align}
Then the total Lagrangian is 
\begin{align}
\mathcal{L}_3 = \mathcal{L}_1 + \mathcal{L}_{{\rm \psi\eta}} + \mathcal{L}_{\eta_{\rm self}}, 
\end{align}
with the mass matrix for each baryon calculated as
\begin{align}
\label{eq_mass_1}
\begin{aligned}
& \hat{M}_N = 
\begin{pmatrix}
g_\psi \alpha & -\sqrt{\frac{3}{2}}y\alpha \\
-\sqrt{\frac{3}{2}}y\alpha & \frac{1}{2}g_\eta \alpha \\
\end{pmatrix}
\ , 
\\
& \hat{M}_\Sigma = 
\begin{pmatrix}
g_\psi \gamma & -\frac{1}{\sqrt{6}}y (\gamma + 2 \alpha)  \\
-\frac{1}{\sqrt{6}}y (\gamma + 2 \alpha)  & - \frac{1}{6} g_\eta (\gamma - 4 \alpha) \\
\end{pmatrix}
\ , \\
& \hat{M}_\Xi = 
\begin{pmatrix}
g_\psi \alpha & -\frac{1}{\sqrt{6}}y (2\gamma +  \alpha) \\
 -\frac{1}{\sqrt{6}}y (2\gamma +  \alpha) & \frac{1}{6} g_\eta (4\gamma - \alpha) \\
\end{pmatrix}, \\
& \hat{M}_\Lambda = 
\begin{pmatrix}
\frac{1}{3}g_\psi(4\alpha - \gamma) & -\frac{1}{6}y(\gamma + 2\alpha) \\
-\frac{1}{6}y(\gamma + 2\alpha) & \frac{1}{2}g_\eta \gamma \\
\end{pmatrix}
\ .
\end{aligned}
\end{align}
Here the components of 2 $\times$ 2 matrices are for $\psi$ and $\eta$ fields. To isolate the contributions from different chiral representations, we first set the mixing term $y=0$. This allows us to determine the mass spectrum arising purely from each representation. Figure~\ref{fig_33bar_36} shows the resulting phenomenological mass spectrum correspond to the case with only the $(3, 6) + (6, 3)$ representation.The ground-state baryons exhibit the mass ordering $m_{\Xi} > m_{\Lambda} > m_{N} > m_{\Sigma}$.

We also note that in this simple model with only $\psi$ and $\eta$ fields, recovering the correct mass ordering between $N$, $\Sigma$, and $\Xi$ requires that the mixing between these two fields should not be large.
To see this, we first express the mass matrices:
\begin{align}
\hat{M}_i = 
\begin{pmatrix}
m_i^{\psi} & \Delta m_i \\
\Delta m_i & m_i^\eta
\end{pmatrix}, \quad i = N, \Sigma, \Xi
\end{align}
These matrix elements correspond to those given in Eq.~(\ref{eq_mass_1}).

Diagonalization yields the physical ground-state masses:
\begin{align}\label{eq_g_s}
m^{\rm G.S.}_i = \frac{1}{2} \left[ (m_i^{\psi} + m_i^\eta ) - \sqrt{ (m_i^\psi - m_i^\eta)^2 + 4 \Delta m_i^2 } \right],
\end{align}
with $ i = N, \Sigma, \Xi$. Since the $\eta$ field contains flavor-symmetric ``bad'' diquarks, we naturally expect $m_i^{\eta} > m_i^\psi$. Combined with the mass hierarchies from the pure $\psi$ field (Sec.~\ref{sec_3_3}) and pure $\eta$ field (Fig.~\ref{fig_33bar_36}), we obtain:
\begin{align}\label{eq_mass_order}
m_\Xi^\eta > m_N^\eta > m_\Sigma^\eta > m_\Sigma^\psi > m_N^\psi = m_\Xi^\psi.
\end{align}
Without mixing ($\Delta m_i = 0$), the ground states would simply be $m_i^{\psi}$. However, nonzero mixing shifts these masses downward. For small mixing (small $y$ and hence small $\Delta m_i$), the ground-state mass can be approximated as:
\begin{align}
m_i^{\rm G.S.} \approx m_i^{\psi} - \frac{(\Delta m_i)^2}{m_i^{\eta} - m_{i}^{\psi}}.
\end{align}
The key observation is that the mass shift depends inversely on the mass gap $(m_i^{\eta} - m_{i}^{\psi})$. From Eq.~(\ref{eq_mass_order}), we have:
\begin{align}
m_\Xi^\eta - m_\Xi^\psi > m_N^\eta - m_N^\psi > m_\Sigma^\eta - m_\Sigma^\psi.
\end{align}
This hierarchy of mass gaps has important consequences. Since the $\Sigma$ has the smallest gap, it experiences the largest downward shift from mixing. The nucleon has an intermediate gap and thus an intermediate shift, while the $\Xi$ has the largest gap and therefore the smallest shift. 

Starting from the degenerate masses $m_N^{\psi} = m_\Xi^{\psi}$ in the pure $\psi$ case, this differential shifting breaks the degeneracy: the $\Sigma$ mass decreases most rapidly, the nucleon mass decreases moderately, and the $\Xi$ mass decreases least. Consequently, for appropriate values of the mixing parameter $y$ (and hence $\Delta m_i$), we obtain $m_\Xi > m_N$. Furthermore, by adjusting the $\eta$ field couplings (such as $g_\eta$) to control the mass gaps $m_i^{\eta} - m_{i}^{\psi}$, we can fine-tune the relative shifts to achieve the ordering: $m_\Xi > m_\Sigma > m_N$.

\subsection{Relation between different mass matrices}

In this subsection, we derive model-independent relations between the traces of different mass matrices that emerge from the chiral structure of our Lagrangian. These relations provide important clue on the baryon mass spectrum.

Using Eq.~(\ref{eq_mass_1}), we calculate the traces of mass matrices for each baryon species:
\begin{equation}\label{eq_trace_form}
\begin{aligned}
{\rm tr}( M_N) &= g_\psi \alpha  + \frac{1}{2} g_\eta \alpha,\\
{\rm tr}( M_\Sigma) &= g_\psi  \gamma   - \frac{1}{6} g_\eta \left(4 \alpha  - \gamma  \right) , \\
{\rm tr}( M_\Xi) &= g_\psi \alpha  + \frac{1}{6} g_\eta \left(4\gamma  - \alpha\right), \\
{\rm tr}( M_\Lambda) &= \frac{1}{3} g_\psi \left( 4\alpha- \gamma  \right) + \frac{1}{2}g_\eta \gamma 
\end{aligned}
\end{equation}
where $\alpha$ and $\gamma$ are the VEVs of the meson field given in Table~\ref{tab-condensate-input}.  These traces exhibit an important symmetry property; in the limit of exact $SU(3)$ flavor symmetry where $\alpha = \beta = \gamma$, all traces become identical: ${\rm tr}( M_N) = {\rm tr}( M_\Lambda)={\rm tr}( M_\Sigma) = {\rm tr}( M_\Xi)$, as expected from flavor symmetry.
Also, we can derive a non-trivial relation by examining the sums of traces. Computing the relevant combinations:
\begin{align}
&{\rm tr}( M_N) + {\rm tr}( M_\Lambda) \nonumber\\
&= \frac{1}{3}\left( 7\alpha  - \gamma\right) g_\psi + \frac{1}{2}\left( \alpha  + \gamma \right) g_\eta,\\
&{\rm tr}( M_\Sigma) + {\rm tr}( M_\Xi) \nonumber\\
&=(\alpha + \gamma ) g_\psi  + \frac{1}{2}(\alpha + \gamma ) g_\eta.
\end{align}
 This leads to the following relation:
\begin{align}
&{\rm tr}( M_N) + {\rm tr}( M_\Lambda) - {\rm tr}( M_\Sigma) - {\rm tr}( M_\Xi) \nonumber\\
&= \frac{4}{3}\left(\alpha - \gamma \right) g_\psi \label{eq_relation}
\end{align}
The difference between these sums depends only on the diagonal $g_\psi$ with $g_\eta$ dependence canceled out completely. It is also noteworthy that this relation holds even when incorporating other chiral representations, such as the (8, 1) + (1, 8) representation, as there are no self-interaction terms for (8, 1) + (1, 8) representation and they do not contribute to the diagonal terms. Consequently, at the leading order of $M$, Eq.~(\ref{eq_relation}) emerges as the most general expression, applicable to scenarios involving 
\begin{enumerate}
 \item Only $(3, \bar{3}) + (\bar{3}, 3)$, 
 \item $(3, \bar{3}) + (\bar{3}, 3)$ and $ (3, 6) + (6, 3)$, 
 \item $(3, \bar{3}) + (\bar{3}, 3) + (3, 6) + (6, 3)$ and $(8, 1) + (1, 8)$ \\(All the representations are included). 
 \end{enumerate}
 In all these cases, the form of Eq.~(\ref{eq_relation}) holds model independent manner. 
 
We can verify this relation using experimental ground-state masses. For the simplest case with only the $(3, \bar{3}) + (\bar{3}, 3)$ representation (1×1 mass matrices), by using the VEV in Table.~\ref{tab-condensate-input} and experimental values:
 \begin{equation}
 \label{eq_gs_input}
 \begin{aligned}
 m_N^{\rm G. S} &= 939 \,{\rm MeV} ,& \quad m_\Lambda^{\rm G. S}&= 1116 \, {\rm MeV}, \\
 m_\Sigma^{\rm G. S} &=1193 \,{\rm MeV} ,& \quad m_\Xi^{\rm G. S} &= 1318  \,{\rm MeV}.
 \end{aligned}
 \end{equation}
We estimate $g_\psi \approx m_N^{\rm G.S}/\alpha = 10.09$. We then obtain the left hand side of Eq.~(\ref{eq_relation}) is 
\begin{align}\label{eq_left}
 m_N^{\rm G. S} + m_\Lambda^{\rm G. S} - m_\Sigma^{\rm G. S} - m_\Xi^{\rm G. S} = -456 \, {\rm MeV}.
\end{align}
while the right hand side of Eq.~(\ref{eq_relation}) is 
\begin{align}\label{eq_right}
\frac{4}{3} (\alpha - \gamma) \frac{m_N^{\rm G. S}}{\alpha}= -457 \, {\rm MeV}
\end{align}
The  agreement validates our theoretical framework. The choice of $g_\psi$ based on the nucleon mass is physically motivated since the $(3, \bar{3}) + (\bar{3}, 3)$ representation, containing "good" diquark configurations, is most directly associated with the nucleon. Here we determine the value of $g_\psi$ from the nucleon ground-state is not an aribitrary choice, but because $(3, \bar{3})+ (\bar{3}, 3)$ representation is most directly tied to the nucleon considering its ``good'' diquark configuration. Although pure $(3, \bar{3}) + (\bar{3}, 3)$ representation itself cannot reproduce the correct mass ordering between $N, \Sigma$ and $\Xi$, Eq.~(\ref{eq_relation}) is still well-satisfied. For case 2 and case 3, the trace of the all the mass matrix remain the form in Eq.~(\ref{eq_trace_form}). The exact value of $g_\psi$ should be obtained in the self-consistent model to reproduce the baryon mass spectrum. 

From Eq.~(\ref{eq_trace_form}), we can derive additional parameter-independent ratios
\begin{align}
\frac{ {\rm tr}(M_\Sigma) + {\rm tr}(M_\Xi) }{{\rm tr}(M_N)} &= \frac{\gamma}{\alpha} + 1 , \label{eq_ratio_1}\\
\frac{ {\rm tr}(M_\Sigma) + {\rm tr}(M_\Lambda) }{{\rm tr}(M_N)} &=\frac{2}{3} \left(\frac{\gamma}{\alpha} + 2 \right), \label{eq_ratio_2}\\
\frac{ {\rm tr}(M_\Xi) - {\rm tr}(M_\Lambda) }{{\rm tr}(M_N)} &=\frac{1}{3} \left(\frac{\gamma}{\alpha} -1 \right). \label{eq_ratio_3}
\end{align}
This should be the general relation for the linear chiral model which includes all the chiral representations. These ratios depend only on $\gamma/\alpha$ and from the ground-state input in Eq.~(\ref{eq_gs_input}), we obtain the ratio the $\gamma / \alpha$
\begin{align}
(\ref{eq_ratio_1}) \Rightarrow \frac{\gamma}{\alpha} = 1.67, \\
(\ref{eq_ratio_2}) \Rightarrow \frac{\gamma}{\alpha} = 1.68, \\
(\ref{eq_ratio_3}) \Rightarrow \frac{\gamma}{\alpha} = 1.65.
\end{align}
We can find that the experimentally-required value $\gamma/\alpha \approx 1.66$ significantly exceeds the standard estimate $\gamma/\alpha = 127/93 = 1.37$ based on pion and kaon decay constants. This 20$\%$ discrepancy cannot be attributed to experimental uncertainties and indicates additional mechanisms such as the explicit chiral symmetry breaking through bare quark masses, which we introduce in the next section.

\section{$SU(3)$ Parity doublet structure}\label{sec_4}
Considering all the above reasons, in the $SU(3)$ case, we argue that the models with chiral representation ($3, \bar{3}$) + ($\bar{3}, 3$) together with (3, 6) + (6, 3) should be the simplest and physical motivated model.
We introduce the following fields:
\begin{equation}
\begin{aligned}
\psi_{1L} \sim (3, \bar{3})_{+1}, \quad &\psi_{1R} \sim (\bar{3}, 3)_{-1}, \\
\psi_{2L} \sim (\bar{3}, 3)_{-1}, \quad &\psi_{2R} \sim (3, \bar{3})_{+1}, \\
\eta_{1L} \sim (3, 6)_{+1}, \quad &\eta_{1R} \sim (6, 3)_{-1}, \\
\eta_{2L} \sim (6, 3)_{-1}, \quad &\eta_{2R} \sim (3, 6)_{+1}.
\end{aligned}
\end{equation}
The fields $\psi_1$ and $\psi_2$ (similarly $\eta_1$ and $\eta_2$) form chiral partners with opposite assignments of the chiral representations---a ``mirror'' assignment. 


Under chiral transformation, these fields transform as
\begin{equation}
\begin{aligned}
\psi_{1L} \rightarrow g_L \psi_{1L} g_{R}^{\dagger}, \quad \psi_{1R} \rightarrow g_{R}\psi_{1R} g_{L}^{\dagger},\\
\psi_{2L} \rightarrow g_{R}\psi_{2L} g_{L}^{\dagger}, \quad \psi_{2R} \rightarrow g_L \psi_{2R} g_{R}^{\dagger},\\
\eta_{1L} \rightarrow g_L \eta_{1L} g_{R}^{\dagger}, \quad \eta_{1R} \rightarrow g_{R}\eta_{1R} g_{L}^{\dagger},\\
\eta_{2L} \rightarrow g_{R}\eta_{2L} g_{L}^{\dagger}, \quad \eta_{2R} \rightarrow g_L \eta_{2R} g_{R}^{\dagger},
\end{aligned}
\end{equation}

We then construct effective Lagrangian based on $SU(3)_L \times SU(3)_R \times U(1)_A$ symmetry as
\begin{align}
\mathcal{L}_0 = \mathcal{L}_{{\rm kin}} + \mathcal{L}_{{\rm CIM}} + \mathcal{L}_{{\rm diag}} + \mathcal{L}_{{\rm off-diag}}
\end{align}
The kinetic term is 
\begin{align}
\mathcal{L}_{{\rm kin}} =& {\rm tr}(\bar{\psi}_{1L}i \slashed{D} \psi_{1L}) + {\rm tr}(\bar{\psi}_{1R}i \slashed{D} \psi_{1R}) \nonumber\\
&+{\rm tr}(\bar{\psi}_{2L}i \slashed{D} \psi_{2L}) + {\rm tr}(\bar{\psi}_{2R}i \slashed{D} \psi_{2R}) \\
& + (\bar{\eta}_{1L})_{(a, \alpha \beta)} i \slashed{D} (\eta_{1L})^{(a, \alpha \beta)}\nonumber \\
&+ (\bar{\eta}_{1R})_{(a b, \alpha)} i \slashed{D} (\eta_{1R})^{(a b, \alpha)} \\
&+(\bar{\eta}_{2L})_{(a b, \alpha)} i \slashed{D} (\eta_{2L})^{(a b , \alpha)}\nonumber \\
&+(\bar{\eta}_{2R})_{(a, \alpha \beta)} i \slashed{D} (\eta_{2R})^{(a , \alpha \beta)}
\end{align}
 with the covariant derivatives for each field are 
 \begin{equation}
 \begin{aligned}
 D_{\mu}\psi_{1L, 2R} =& \partial_\mu \psi_{1L, 2R} - i \mathcal{L}_\mu \psi_{1L, 2R} + i \psi_{1L, 2R}\mathcal{R}_\mu, \\
 D_{\mu}\psi_{1R, 2L} =& \partial_\mu \psi_{1R, 2L} - i \mathcal{R}_\mu \psi_{1R, 2L} + i \psi_{1R, 2L}\mathcal{L}_\mu, \\
 (D_\mu \eta_{1L, 2R})^{(a, \alpha \beta)} =& \partial_\mu \eta_{1L, 2R}^{(a, \alpha \beta)} - i (\mathcal{L}_\mu)^a_b \eta_{1L, 2R}^{(b, \alpha \beta)} \\
 &- i \left[ (\mathcal{R}_\mu)^{\alpha}_{\rho}\delta_\sigma^\beta + \delta_\rho^\alpha (\mathcal{R}_\mu)^\beta_\sigma \right] \eta_{1L, 2R}^{(a, \rho \sigma)}, \\
  (D_\mu \eta_{1R, 2L})^{(a b , \alpha)} =& \partial_\mu \eta_{1R, 2L}^{(a b, \alpha)} - i (\mathcal{R}_\mu)^{\alpha}_\beta \eta_{1R, 2L}^{(a b, \beta)} \\
 &- i \left[ (\mathcal{L}_\mu)^{a}_{c}\delta^b_d + \delta^a_c (\mathcal{L}_\mu)^b_d \right] \eta_{1R, 2L}^{(c d, \alpha)}, 
 \end{aligned}
 \end{equation}
 where the $\mathcal{L}_\mu$ and $\mathcal{R}_\mu$ represent the external gauge fields arising from the gauging of the chiral $SU(3)_L \times SU(3)_R$ symmetry.
The chiral invariant mass term is expressed as
\begin{align}
\mathcal{L}_{{\rm CIM}} =&  m_0^{(1)} \left[ {\rm tr}(\bar{\psi}_{1R}  \psi_{2L}) + {\rm tr}(\bar{\psi}_{1L} \psi_{2R})\right] \nonumber\\
&+ m_0^{(2)}\left[ {\rm tr} (\bar{\eta}_{1L}\eta_{2R}) + {\rm tr} ( \bar{\eta}_{1R}\eta_{2L} ) \right]
\end{align}
The diagonal Yukawa interactions are given by
\begin{align}
&\lag_\mathrm{diag}= \notag \\
&\quad g_1\left[
\varepsilon_{r_1r_2r_3}\varepsilon^{l_1l_2l_3}
(\bar\psi_{1L})^{r_1}_{l_1}(M^\dag)^{r_2}_{l_2}(\psi_{1R})^{r_3}_{l_3}
+\mathrm{h.c.}\right] \nonumber \\
&+g_2\left[
\varepsilon_{l_1l_2l_3}\varepsilon^{r_1r_2r_3}
(\bar\psi_{2L})^{l_1}_{r_1}(M)^{l_2}_{r_2}(\psi_{2R})^{l_3}_{r_3}
+\mathrm{h.c.}\right] \label{eq_g1g2}\\
& + g_3 \left[ (\bar{\eta}_{1R})_{(ab, \alpha)} (M)^a_\beta (\eta_{1L})^{ (b, \alpha \beta)} + {\rm h. c.} \right] \nonumber\\
& + g_4 \left[ (\bar{\eta}_{2R})_{(a, \alpha\beta)} (M^{\dagger})^\alpha_b (\eta_{2L})^{ (a b, \beta)} + {\rm h. c.} \right]
\end{align}
The interaction between different chiral representations (off-diagonal) of baryons are 
\begin{align}
&\mathcal{L}_{{\rm off-diag}} = \notag \\
& \quad y_1 \left[\epsilon_{abc} (\bar{\psi}_{1R})^a_\alpha (M)_\beta^b (\eta_{1L})^{(c, \alpha \beta)} + {\rm h.c.} \right.  \nonumber\\
&\quad \left.  \epsilon_{\alpha \beta \sigma} (\bar{\psi}_{1L})^\alpha_a (M^\dagger)^\beta_b (\eta_{1R})^{(ab, \sigma)}    + {\rm h.c.}  \right]\\
& +y_2 \left[\epsilon_{\alpha \beta \sigma} (\bar{\psi}_{2R})^\alpha_a (M^{\dagger})_b^\beta (\eta_{2L})^{(a b , \sigma)} + {\rm h.c.} \right.  \nonumber\\
&\quad \left.  \epsilon_{a b c} (\bar{\psi}_{2L})^a_\alpha (M)^b_\beta (\eta_{2R})^{(c, \alpha \beta)}    + {\rm h.c.}  \right] \label{eq_y2}
\end{align}
For simplicity, we also define the isospin multiplets as 
\begin{equation}
\begin{aligned}
\psi_N&\equiv(\psi_p,\psi_n)\\
\psi_\Sigma&\equiv(\psi_{\Sigma^-},\psi_{\Sigma^0},\psi_{\Sigma^+})\\
\psi_\Xi&\equiv(\psi_{\Xi^-},\psi_{\Xi^0})\,. 
\end{aligned}
\end{equation}
Due to the mixing of different parity states, we define each baryon field as
\begin{align}
\Psi_i = (\psi_{1i}, \eta_{1i}, \gamma_5 \psi_{2i}, \gamma_5 \eta_{2i})^{T}, (i = N,\Lambda, \Sigma, \Xi),
\end{align}
we can write the mass term as
\begin{align}
\mathcal{L}_{{\rm mass}} = - \sum_{i = N, \Lambda, \Sigma, \Xi} \bar{\Psi}_i \hat{M}_i \Psi_i,
\label{eq_mass_L}
\end{align} 
the corresponding mass matrix for each baryon is calculated as
\begin{align}
&\hat{M}_N  (g_1, g_2, g_3, g_4, y_1, y_2) = \nonumber \\
&\left[\begin{matrix}
g_1 \alpha  & 
-\sqrt{\frac{3}{2}} y_1 \alpha & m_0^{(1)} & 0 \\
-\sqrt{\frac{3}{2}} y_1 \alpha&  \frac{1}{2}g_3\alpha & 0 & m_0^{(2)}\\
m_0^{(1)} & 0 & -g_2 \alpha & \sqrt{ \frac{3}{2} }y_2\alpha \\
0 & m_0^{(2)} & \sqrt{ \frac{3}{2} }y_2\alpha & -\frac{1}{2}g_4 \alpha
\end{matrix} \right] \label{eq_mass_N}\\
&\hat{M}_\Sigma  (g_1, g_2, g_3, g_4, y_1, y_2) =\nonumber \\
&\left[\begin{matrix}
g_1 \gamma  & 
-\frac{1}{\sqrt{6}}y_1(\gamma + 2 \alpha) & m_0^{(1)} & 0\\
 &  -\frac{1}{6}g_3(\gamma - 4 \alpha)
 & 0 & m_0^{(2)}\\
 &  & -g_2 \gamma & \frac{1}{\sqrt{6}}y_2(\gamma + 2 \alpha)\\
 & &  & \frac{1}{6}g_4 (\gamma - 4\alpha)
\end{matrix} \right]\\
&\hat{M}_\Xi (g_1, g_2, g_3, g_4, y_1, y_2)=  \nonumber\\
& \left[\begin{matrix}
g_1 \alpha  & 
-\frac{1}{\sqrt{6}} y_1 (2\gamma + \alpha) & m_0^{(1)} & 0\\
  & \frac{1}{6}g_3(4\gamma - \alpha) & 0 & m_0^{(2)}\\
  & &-g_2 \alpha & \frac{1}{\sqrt{6}}y_2(2\gamma + \alpha) \\
  & & & -\frac{1}{6}g_4(4\gamma - \alpha)
\end{matrix} \right] \\
&\hat{M}_\Lambda (g_1, g_2, g_3, g_4, y_1, y_2) =\nonumber\\
& \left[\begin{matrix}
-\frac{g_1}{3}  (\gamma - 4 \alpha)  & 
 -\frac{y_1}{\sqrt{6}} (\gamma + 2\alpha) & m_0^{(1)} & 0\\
&   \frac{g_3}{2}\gamma & 0 & m_0^{(2)}\\
&   &\frac{g_2}{3}( \gamma - 4 \alpha ) & \frac{y_2}{\sqrt{6}} (\gamma + 2 \alpha)\\
& & & -\frac{g_4}{2} \gamma
\end{matrix} \right] \label{eq_mass_Lam}, 
\end{align}
Also, to account for the mass difference between different baryon chiral representations, it is also necessary to consider effect of the explicit breaking of the chiral symmetry. To do that, we have to introduce the bare mass matrix
\begin{align}
\mathcal{M} \sim (3, \bar{3})_{-2}.
\label{eq_bare_quark_mass}
\end{align}
with the explicit form as
\begin{align}
\langle \mathcal{M} \rangle = 
\left[ 
\begin{matrix}
m_u & 0 & 0\\
0 & m_d & 0\\
0 & 0 & m_s
\end{matrix}
\right].
\end{align}
Considering the isospin symmetry, we set $m_u = m_d = 3. 5$ MeV~\cite{ParticleDataGroup:2024cfk}. From the ratio of the mass of pion and kaon, we can obtain the value of $m_s / m_u$ as
\begin{align}
\frac{m_s + m_u}{m_u + m_d} = \frac{m_{K}^2 }{m_{\pi}^2} = 13.6
\end{align}
where we take the value of $m_\pi = 135$ MeV and $m_K = 497$ MeV. Then we find
\begin{align}
\frac{m_s}{m_u} = 26.1.
\end{align}
The Lagrangian including the bare mass matrix $\mathcal{M}$ share the similar form to Eq.~(\ref{eq_g1g2}) to Eq.~(\ref{eq_y2})
\begin{align}
&\mathcal{L}_{\mathcal{M}} = \notag \\
&\quad \tilde{g}_1\left[
\varepsilon_{r_1r_2r_3}\varepsilon^{l_1l_2l_3}
(\bar\psi_{1L})^{r_1}_{l_1}(\mathcal{M}^\dag)^{r_2}_{l_2}(\psi_{1R})^{r_3}_{l_3}
+\mathrm{h.c.}\right] \nonumber \\
&+\tilde{g}_2\left[
\varepsilon_{l_1l_2l_3}\varepsilon^{r_1r_2r_3}
(\bar\psi_{2L})^{l_1}_{r_1}(\mathcal{M})^{l_2}_{r_2}(\psi_{2R})^{l_3}_{r_3}
+\mathrm{h.c.}\right] \\
& + \tilde{g}_3 \left[ (\bar{\eta}_{1R})_{(ab, \alpha)} (\mathcal{M})^a_\beta (\eta_{1L})^{ (b, \alpha \beta)} + {\rm h. c.} \right] \nonumber\\
& + \tilde{g}_4 \left[ (\bar{\eta}_{2R})_{(a, \alpha\beta)} (\mathcal{M}^{\dagger})^\alpha_b (\eta_{2L})^{ (a b, \beta)} + {\rm h. c.} \right] \\
& +\tilde{y}_1 \left[\epsilon_{abc} (\bar{\psi}_{1R})^a_\alpha (\mathcal{M})_\beta^b (\eta_{1L})^{(c, \alpha \beta)} + {\rm h.c.} \right.  \nonumber\\
&\quad \left.  \epsilon_{\alpha \beta \sigma} (\bar{\psi}_{1L})^\alpha_a (\mathcal{M}^\dagger)^\beta_b (\eta_{1R})^{(ab, \sigma)}    + {\rm h.c.}  \right]\\
&+ \tilde{y}_2 \left[\epsilon_{\alpha \beta \sigma} (\bar{\psi}_{2R})^\alpha_a (\mathcal{M}^{\dagger})_b^\beta (\eta_{2L})^{(a b , \sigma)} + {\rm h.c.} \right.  \nonumber\\
&\quad \left.  \epsilon_{a b c} (\bar{\psi}_{2L})^a_\alpha (\mathcal{M})^b_\beta (\eta_{2R})^{(c, \alpha \beta)}    + {\rm h.c.}  \right] .
\end{align}
There are 12 parameters and to simplify the model, we assume the diagonal coupling $g$ terms and the coupling between different fields $y$ terms have the following relation
\begin{align}
\tilde{g}_i &= z_0 g_i, \quad (i = 1, 2, 3, 4),\\
\tilde{y}_i &= z_1 y_j , \quad (j = 1, 2).
\end{align}
The bare quark mass matrix $\mathcal{M}$ for explicit chiral symmetry breaking and the scalar meson field $M$ are introduced to have the same chiral representation so that the Yukawa couplings involving $\mathcal{M}$ and $M$ should have similar chiral structures and it is reasonable to assume them are related by a common proportionality factor.

Then the whole Lagrangian is 
\begin{align}
\mathcal{L} = \mathcal{L}_0 + \mathcal{L}_{\mathcal{M}}.
\end{align}
with 8 parameters to be determined.

\section{Numerical analysis}\label{sec_5}
In this section, we will perform the numerical analysis to fit for the mass spectra of baryons.
We first review about the Gell-Mann-Okubo mass relation 
\subsection{Brief review of Gell-Mann-Okubo mass relation}
The Gell-Mann-Okubo (GMO) mass relation provides an important guideline to understand  the role of SU(3) breaking of baryons.. To illustrate this, consider a simple effective Lagrangian with flavor $SU(3)$ symmetry:

\begin{align}\label{eq-GO}
\lag^\mathrm{V}=
-a\tr\bar{B}B
-b\tr\bar{B}MB
-c\tr\bar{B}BM\,, 
\end{align}
where $B$ is a $3 \times 3$ matrix containing the octet baryon fields, $M$ represents the meson field matrix, and $a$, $b$, $c$ are real coupling constants. It is important to note that while this Lagrangian respects flavor $SU(3)$ symmetry, it does not possess the full chiral $SU(3)_{ L} \times SU(3)_{ R}$ symmetry.

When the meson field acquires a vacuum expectation value $\langle M \rangle = \mathrm{diag}(\alpha, \beta, \gamma)$ with SU(3) breaking as accommodated by the mass matrix ${\mathcal M}$ in Eq.~(\ref{eq_bare_quark_mass}) but with keeping the isospin symmetry ($\alpha = \beta$), the octet baryon masses become:
\begin{equation}
\begin{aligned}
m_N&=a+b\alpha+c\gamma \ , \\
m_\Sigma&=a+b\alpha+c\alpha\ , \\
m_\Xi&=a+b\gamma+c\alpha\ , \\
m_\Lambda&=a+b\frac{\alpha+2\gamma}{3}+c\frac{\alpha+2\gamma}{3}\ . 
\end{aligned}
\end{equation}
Eliminating the parameters $a$, $b$, and $c$ from these equations yields the following relation: 
\begin{align}
\frac{m_N+m_\Xi}{2}=\frac{3m_\Lambda+m_\Sigma}{4}\,, 
\end{align}
which is called the GMO  mass relation for octet baryons. By eliminating the factor $\gamma / \alpha$ from Eq.~(\ref{eq_ratio_1}) to (\ref{eq_ratio_3}), we obtain an expression identical in form to the GMO  relation:
\begin{align}
\frac{{\rm tr} (M_N) + {\rm tr} (M_\Xi) }{2} = \frac{3 {\rm tr} (M_\Lambda) + {\rm tr} (M_\Sigma)}{4},
\end{align}
which represents the sum of mass that extends to higher excited states. The GMO relation holds at the mass matrix level regardless of which chiral representations are included.

In a naive quark mass counting,
the GMO mass relation is satisfied by assuming 
$M_u \simeq M_d$, $m_N \sim 3M_u$, $m_\Xi \sim M_u + 2M_s$, $m_\Lambda \sim 2M_u + M_s$, and  $m_\Sigma \sim 2M_u + M_s$, 
where $M_q$ ($q=u,d,s$) are the constituent quark masses. 
These estimates hold for typical constituent quark models. 
On the other hand, these quark counting is sufficient but not necessary conditions;
the {\GO} is a weaker condition than that deduced from the quark counting.

\begin{table*}
\caption{
Physical inputs and predicted states for the baryon masses belonging to four $SU(3)$-flavor octets.  
}
\label{tab-mass-inputs}
\centering
\begin{tabular}{c||c|c|c|c}
 & \multicolumn{4}{c||}{Mass inputs for octet members [MeV]} \\
\hline\hline
$J^P$ & $N$ & $\Lambda$ & $\Sigma$ & $\Xi$  \\
\hline
$m_1: 1/2^+$(G.S.) & 
$N(939)$: $939$ & 
$\Lambda(1116)$: $1116$ & 
$\Sigma(1193)$: $1193$ & 
$\Xi(1318)$: $1318$  \\
$m_2: 1/2^+$ & 
$N(1440)$: $1440$ & 
$\Lambda(1600)$: $1600$ & 
$\Sigma(1660)$: $1660$ & 
$\Xi(?)$:    \\
$m_3: 1/2^-$ & 
$N(1535)$: $1530$ & 
$\Lambda(1670)$: 1674 & 
$\Sigma(?)$:  & 
$\Xi(?)$:     \\
$m_4: 1/2^-$ & 
$N(1650)$: $1650$ & 
$\Lambda(?)$:  & 
$\Sigma(?)$:  & 
$\Xi(?)$:     \\
\hline
\end{tabular}
\end{table*}

\subsection{Numerical fitting}

\begin{figure*}\centering
\includegraphics[width=1\hsize]{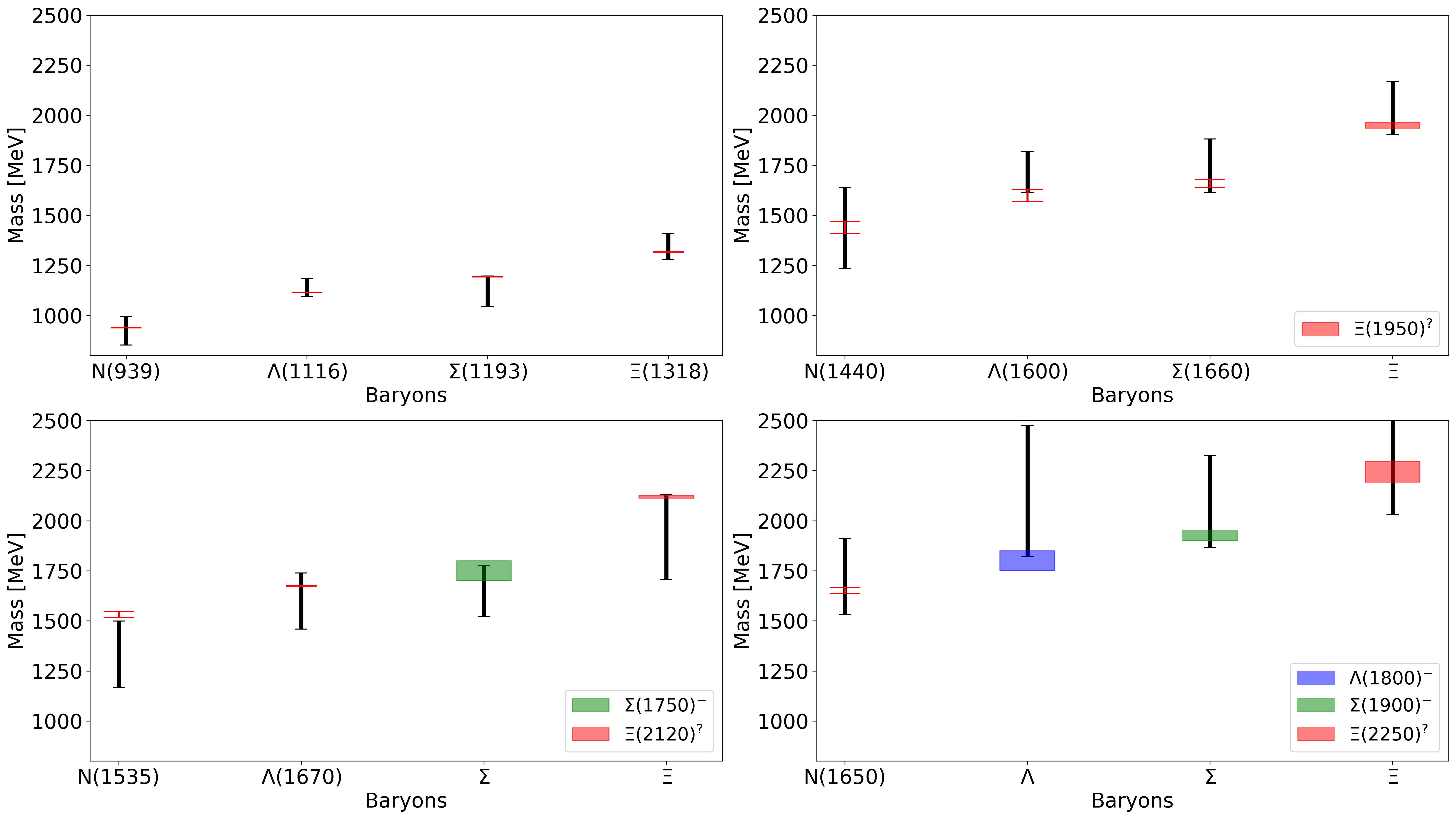}
\caption{Numerical results for fitting the octet baryon masses with the combination of chiral invariant mass $( m_0^{(1)}, m_0^{(2)}) = (800, 1000)$ MeV. The red lines show the experimental values with $\delta m_i$ and the black lines show our numerical results for accumulating spectra from all parameter set satisfying $f_{{\rm min}}<1$. }
\label{fig-massspec}
\end{figure*}
\begin{figure}\centering
\includegraphics[width=1\hsize]{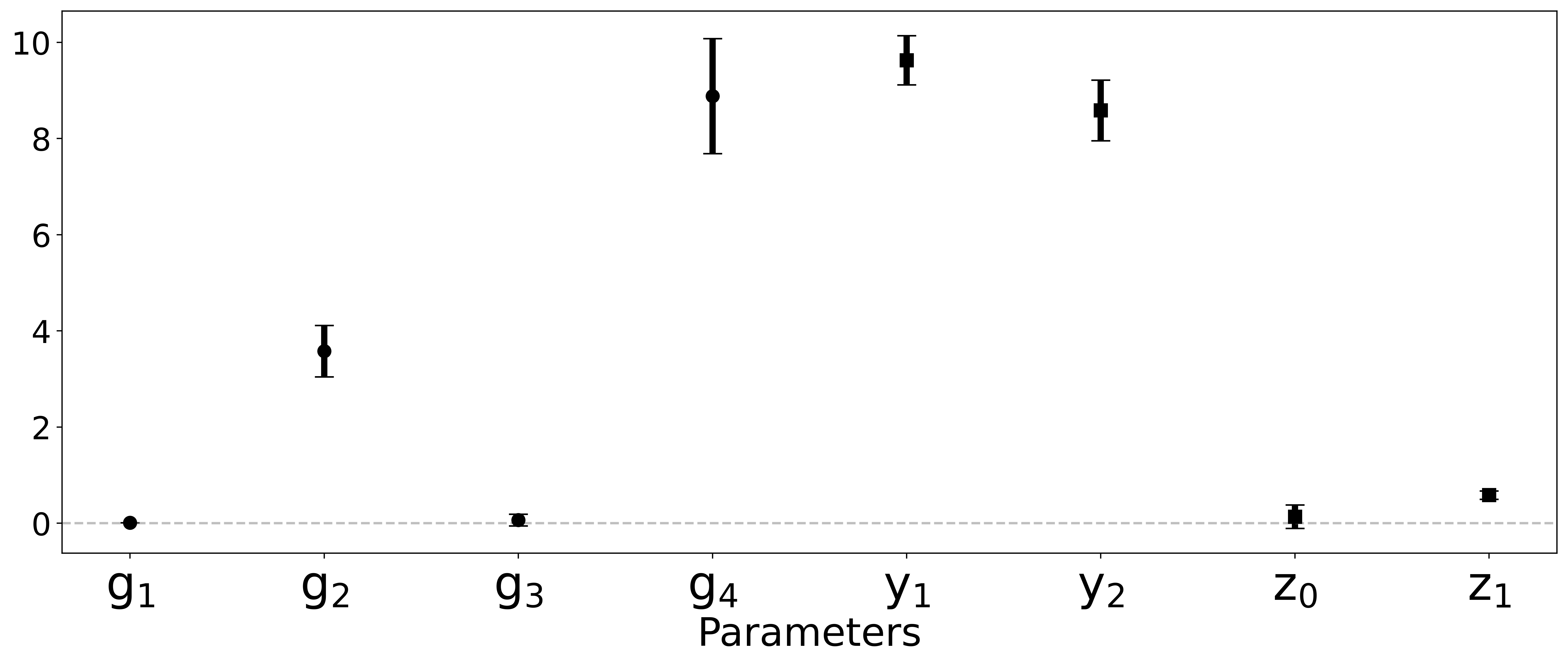}
\caption{Parameters determined from fitting the octet baryon masses with chiral invariant mass $( m_0^{(1)}, m_0^{(2)}) = (800, 1000)$ MeV correspond to Fig.~\ref{fig-massspec}. }
\label{fig-para}
\end{figure}

In this subsection, we numerically fit the model parameters to the known light-baryon masses. 
We determine the eight coupling constants $(g_1, g_2, g_3, g_4, y_1, y_2, z_0, z_1)$ by minimizing the following error function using the 10 mass values listed in Table~\ref{tab-mass-inputs} for a given parameter set of chiral invariant masses $\left(m_0^{(1)}, m_0^{(2)}\right)$:
\begin{align}
f_{\rm min} = \frac{1}{2}\sum_{i=1}^{10} \left( \frac{ m_i^{\rm fitted} - m_i^{\rm input} }{\delta m_i} \right)^2
\end{align}
where $\delta m_i = 0.1 \times m_i^{\rm input}$ represents a 10\% theoretical uncertainty. This uncertainty accounts for higher-order corrections and model approximations not included in our leading-order analysis. The coefficient $1/2$ follows the standard definition of normalized $\chi^2$ statistics.
As discussed in the previous section, the chiral invariant masses must satisfy the hierarchy $m_0^{(2)} > m_0^{(1)}$ to ensure the correct mass ordering between the $(3,6) + (6,3)$ and $(3,\bar{3}) + (\bar{3},3)$ representations.

Figure~\ref{fig-massspec}  shows the resulting mass spectrum for the case of $(m_0^{(1)}, m_0^{(2)}) = (800, 1000)$ MeV. The red lines indicate experimental masses and their uncertainties from the Particle Data Group. The black lines show our theoretical predictions, compiled from all parameter sets that successfully fit the data ($f_{\rm min} < 1$). For excited states where experiments have not yet determined the quantum numbers, we propose tentative spin-parity assignments based on mass proximity to observed resonances—these candidate identifications are marked by colored shaded regions.

Our model shows good agreement with experimental data across all baryon sectors, and the GMO relation is satisfied for both the ground and excited states. In the nucleon sector, both the ground state $N(939)$ and the Roper resonance $N(1440)$ are well reproduced. The negative-parity $N(1535)$ appears slightly below its experimental value, which is consistent with expectations given that this state likely has  meson-baryon molecular structure not captured in our approach~\cite{Kaiser:1995cy, Kaiser:1996js, Inoue:2001ip}. For the $\Lambda$ baryons, the model accurately reproduces the ground state $\Lambda(1116)$, first positive-parity excitation $\Lambda(1600)$  and the first negative-parity excitation $\Lambda(1670)$. Additionally, we predict a second negative-parity state to have the mass from 1800 MeV to 2500 MeV, which we identify with the $\Lambda(1800)$ resonance observed experimentally.
For  $\Sigma$ sector, our model successfully describes both the ground state $\Sigma(1193)$ and the first positive-parity excitation $\Sigma(1660)$. We identify two negative-parity states corresponding to the experimentally observed $\Sigma(1750)$ and $\Sigma(1900)$ resonances. 

\begin{figure*}\centering
\includegraphics[width=1\hsize]{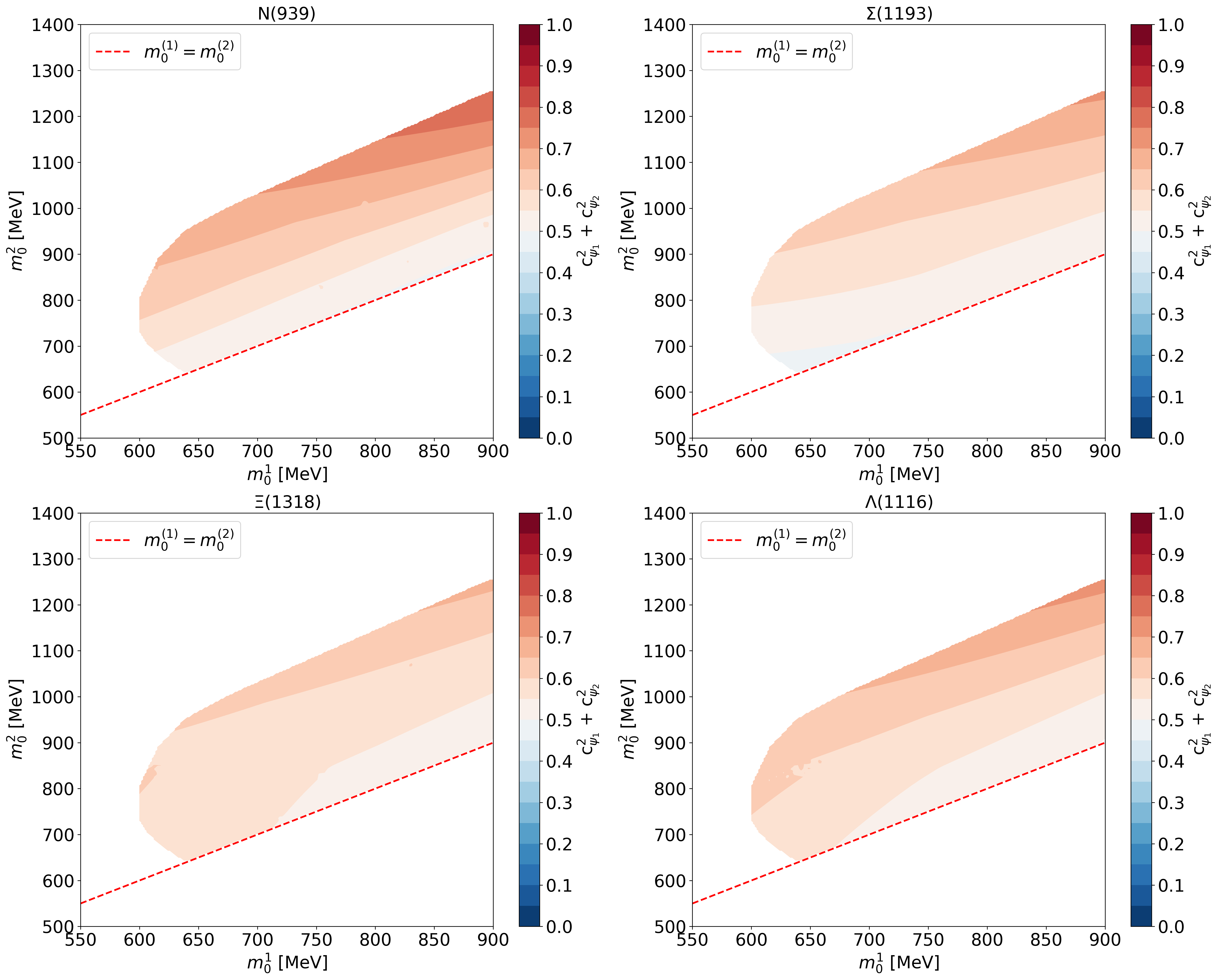}
\caption{Numerical results for the (3, $\bar{3}$) + ($\bar{3}$, 3) composition of the ground state $N(939), \Sigma(1193), \Xi(1318)$ and $\Lambda(1116)$ for different choices of $(m_0^{(1)}, m_0^{(2)})$. The colored region show where we can find a good solution $f_{{\rm min}}^{{\rm best}}<1$ and the region without color indicate that for such combination of $(m_0^{(1)}, m_0^{(2)})$, the $f_{{\rm min}}^{{\rm best}}$ is larger than 1. The colored values are obtained from $c_{\psi_1}^2 + c_{\psi_2}^2$.  }
\label{fig-groundratio}
\end{figure*}
\begin{figure*}\centering
\includegraphics[width=1\hsize]{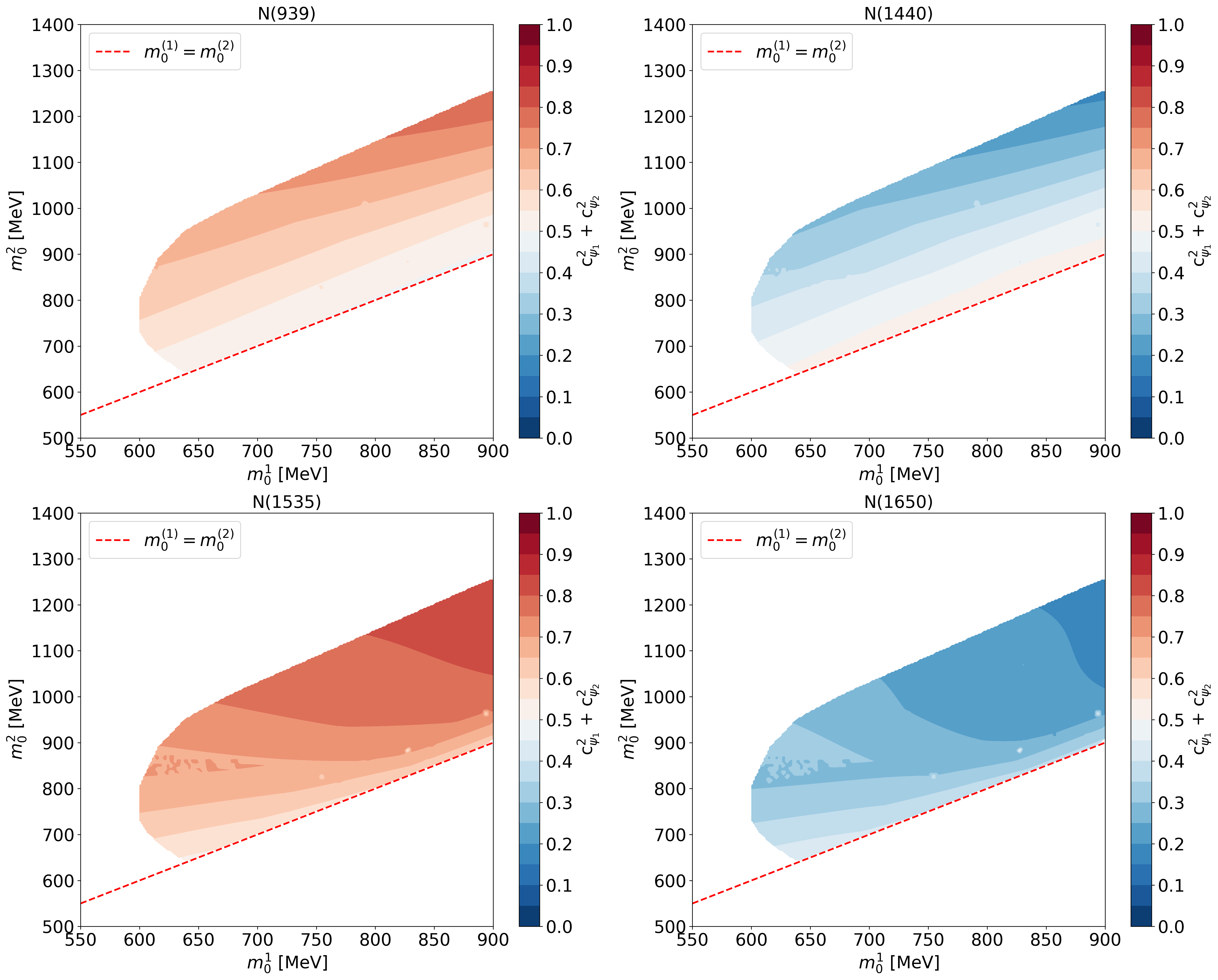}
\caption{Numerical results for the (3, $\bar{3}$) + ($\bar{3}$, 3) composition of four nucleon state $N(939), N(1440), N(1535)$ and $N(1650)$ for different choices of $(m_0^{(1)}, m_0^{(2)})$. The colored region show where we can find a good solution $f_{{\rm min}}^{{\rm best}}<1$ and the region without color indicate that for such combination of $(m_0^{(1)}, m_0^{(2)})$, the $f_{{\rm min}}^{{\rm best}}$ is larger than 1. The colored values are obtained from $c_{\psi_1}^2 + c_{\psi_2}^2$.  }
\label{fig-mix_N}
\end{figure*}

The $\Xi$ sector presents unique challenges due to limited experimental data on excited states. Below 2 GeV, several $\Xi$ resonances have been observed---$\Xi(1530)$, $\Xi(1620)$, $\Xi(1690)$, $\Xi(1820)$, and $\Xi(1950)$---but their quantum numbers remain undetermined. Only $\Xi(1530)$ and $\Xi(1820)$ have well-established spin assignments of $J = 3/2$ with three-star ratings in the PDG~\cite{ParticleDataGroup:2024cfk}. The states $\Xi(1620)$ and $\Xi(1690)$, whose quantum numbers remain experimentally undetermined, have been suggested by multiple theoretical studies to possess negative parity~\cite{Ramos:2002xh, Oh:2007cr,Huang:2020taj,Sekihara:2015qqa,Sekihara:2016hcf}.
For positive-parity excitations, we predict the first $J^P = 1/2^+$ excited state at approximately 2000 MeV, which could be assigned as $\Xi(1950)$. This assignment is consistent with several theoretical studies that suggest $\Xi(1950)$ carries spin-1/2 and positive parity~\cite{Oh:2007cr,Xiao:2013xi}. 
 For negative-parity excitations, based on our predicted masses,  we tentatively assign as $\Xi(2120)$ and $\Xi(2250)$, respectively. These predictions provide valuable guidance for future experimental searches and spin-parity determinations in the $\Xi$ spectrum. Table~\ref{tab-mass-predict} summarizes our results for the octet baryon spectrum, including both the input values (marked with $*$) and the predictions. 
 
Figure~\ref{fig-para} shows the determined parameters from our fit. The Yukawa coupling constants are of order $\mathcal{O}(1-10)$, consistent with our previous two-flavor analysis discussed in Sec.~\ref{sec_su2PDM}. 
We find that the self-coupling constants $g_1$ and $g_3$ (corresponding to the $\psi_1$ and $\eta_1$ fields with $(3,\bar{3})$ and $(3,6)$ representations) are nearly vanishing, while $g_2$ and $g_4$ (for the mirror fields $\psi_2$ and $\eta_2$) take much larger values. Additionally, the Yukawa couplings $y_1$ and $y_2$ are relatively large, indicating significant mixing between different chiral representations.

\begin{table*}
\caption{
Physical inputs (marked with $*$) and predicted states for baryon masses belonging to four $SU(3)$-flavor octets.  
}
\label{tab-mass-predict}
\centering
\begin{tabular}{c||c|c|c|c|c}
 & \multicolumn{4}{c||}{Mass inputs for octet members [MeV]} \\
\hline\hline
$J^P$ & $N$ & $\Lambda$ & $\Sigma$ & $\Xi$ & $\frac{m_{N}+m_{\Xi}}{2} - \frac{3m_{\Lambda} + m_{\Sigma}}{4}$ \\
\hline
$m_1: 1/2^+$(G.S.) & 
$N(939)$: $939^*$ & 
$\Lambda(1116)$: $1116^*$ & 
$\Sigma(1193)$: $1193^*$ & 
$\Xi(1318)$: $1318^*$  & - 6.75 \, {\rm MeV} \\
$m_2: 1/2^+$ & 
$N(1440)$: $1440^*$ & 
$\Lambda(1600)$: $1600^*$ & 
$\Sigma(1660)$: $1660^*$ & 
$\Xi(1950)$: 1950  ~& 80 \, {\rm MeV} \\
$m_3: 1/2^-$ & 
$N(1535)$: $1530^*$ & 
$\Lambda(1670)$: 1674$^*$ & 
$\Sigma(1750)$: 1750 ~& 
$\Xi(2120)$: 2120 & 132 \, {\rm MeV}   \\
$m_4: 1/2^-$ & 
$N(1650)$: $1650^*$ & 
$ \Lambda(1800)$: 1800 ~& 
$\Sigma(1900)$: 1925 ~& 
$\Xi(2250)$: 2250 ~& 118.75 \, {\rm MeV}\\
\hline
\end{tabular}
\end{table*}

We next examine the composition of baryon mass eigenstates in terms of the different chiral representations. Each physical baryon state can be expressed as a linear combination of the basis fields:
\begin{align}
| N \rangle = c_{\psi_1} | \psi_1 \rangle + c_{\eta_1} | \eta_1 \rangle + c_{\psi_2} | \psi_2 \rangle + c_{\eta_2} | \eta_2 \rangle
\end{align}
where the coefficients satisfy the normalization condition $c_{\psi_1}^2 + c_{\eta_1}^2 + c_{\psi_2}^2 + c_{\eta_2}^2 = 1$. Figure~\ref{fig-groundratio} displays the fraction of the $(3,\bar{3}) + (\bar{3},3)$ representation (quantified by $c_{\psi_1}^2 + c_{\psi_2}^2$) in the ground-state baryons as a function of the chiral invariant masses $(m_0^{(1)}, m_0^{(2)})$. The uncolored regions correspond to parameter choices where $f_{\rm min}^{\rm best} > 1$, indicating poor fits to the experimental data. First, we find that to have a good fit ($f_{\rm min}^{\rm best} \leq 1$) requires both chiral invariant masses to be sufficiently large. Specifically, $m_0^{(1)}$ must exceed 600 MeV, consistent with constraints from neutron star observations in the SU(2) parity doublet model \cite{Marczenko:2018jui,Minamikawa:2020jfj,Minamikawa:2023eky,Kong:2023nue,Gao:2024chh,Kong:2025dwl,Gao:2025nkg,Gao:2025vdc}. Similarly, $m_0^{(2)}$ must lie within the range 650--1250 MeV, in agreement with our previous studies. Second, across all successful parameter regions, the $(3,\bar{3}) + (\bar{3},3)$ representation consistently dominates the ground-state composition, with $c_{\psi_1}^2 + c_{\psi_2}^2 > 0.5$. This dominance becomes even more pronounced as $m_0^{(2)}$ increases for fixed $m_0^{(1)}$, reflecting the physical expectation that the ``bad'' diquark configurations in the $(3,6) + (6,3)$ representation are energetically suppressed.

We also show the $\psi$ field ratio for different nucleon state with different choices of $(m_0^{(1)}, m_0^{(2)})$ as in Fig.~\ref{fig-mix_N}. We find that for the ground state nucleon and its first-excited state with negative parity $N(1535)$, $\psi$ fields are dominant. While for $N(1440)$ (first excited state with positive parity) and $N(1650)$ (Second excited state with negative parity) are dominated by the $\eta$ field. With the increasing of the $m_0^{(2)}$ value, the $\psi$ field ratio become larger in $N(939)$ and $N(1535)$ while in $N(1440)$ and $N(1650)$, the $\psi$ field ratio become smaller. We have to note that $N(939)$ and $N(1535)$ are regarded as chiral partners; $N(1440)$ and $N(1650)$ are chiral partners. The $N(939)$ is the lowest state with positive parity and $N(1535)$ is the lowest state with negative parity. For $N(939)$ and $N(1535)$, $\psi$ field is dominated. Since the eigenvectors are orthogonal,  then the $\psi$ ratio for $N(1440)$ and $N(1650)$ is small. This feature applies also to other baryon species such as $\Lambda, \Sigma$ and $\Xi$.


\section{Summary and Discussion}\label{sec-summary}

 In this work, we have developed a linear $SU(3)_L \times SU(3)_R$ parity doublet model that successfully describes the octet baryon mass spectrum. In the context of SU(2) chiral symmetry, Ref.~\cite{Jido:1999hd} proposed a quartet scheme where $\Delta_\pm$ and $N^*_\pm$ with even and odd parity form a chiral multiplet in the $(1, \frac{1}{2}) \oplus (\frac{1}{2}, 1)$ representation. Using the mirror assignment, they derived parameter-free mass relations among the quartet members, which were shown to be well satisfied by observed resonances with spin $J = \frac{1}{2}, \frac{3}{2}, \frac{5}{2}$. Their work demonstrated that chiral symmetry provides meaningful constraints on excited baryon spectra and established the theoretical foundation for studying parity doubling beyond the ground-state nucleon. The present work extends this approach to the three-flavor sector.

Through systematic analysis, we demonstrated why specific chiral representations are necessary for reproducing the correct baryon mass hierarchy. Starting from the minimal $(3,\bar{3}) + (\bar{3},3)$ representation alone, we showed that this leads to unphysical degeneracies in the mass spectrum. While adding the $(8,1) + (1,8)$ representation resolves these degeneracies, it introduces incorrect mass orderings. This analysis establishes that the $(3,6) + (6,3)$ representation, despite containing energetically disfavored ``bad'' diquarks, is essential for achieving the correct mass spectrum---particularly the $\Sigma$-$\Xi$ mass ordering. Our simplified model excludes the $(8,1) + (1,8)$ representation, reducing the number of parameters while retaining the essential physics. This choice is motivated by theoretical considerations: the $(8,1)$ representation lacks clear diquark-quark separation and cannot be constructed from standard $SU(6)$ wave functions (see additional discussion in Appendix~\ref{append_1}).

We also incorporated explicit chiral symmetry breaking through quark mass terms, which was absent in previous analyses~\cite{Minamikawa:2023ypn, Gao:2024mew}. This allows us to properly account for $SU(3)$ flavor breaking effects and study the interplay between spontaneous and explicit symmetry breaking. Our analysis reveals that flavor symmetry breaking from different meson condensates dominates over explicit chiral symmetry breaking from quark mass differences.

Our numerical results successfully reproduce the ground-state octet baryon masses and predict a rich spectrum of excited states. The model accurately describes well-established resonances including $N(1440)$, $N(1535)$, $\Lambda(1600)$, $\Lambda(1670)$, and $\Sigma(1660)$. For less well-determined states, particularly in the $\Xi$ sector where experimental data is limited, we  predict the first positive-parity excitation $\Xi(1950)$ and negative-parity states that we identify with $\Xi(2120)$, and $\Xi(2250)$. 

The analysis of baryon composition reveals that ground states are dominated by the $(3,\bar{3}) + (\bar{3},3)$ representation (typically $>50\%$), consistent with the physical expectation that ``good'' diquark configurations are energetically favored. However, the $(3,6) + (6,3)$ contribution remains crucial for generating the correct mass hierarchy and contributes significantly to excited states. 

Finally, we note that our calculation yields axial charge $g_A \approx 0.1$, significantly smaller than the experimental value $g_A^{\exp} = 1.27$. This suppression arises from the substantial mixing between the original fields and their mirror partners, required by the large chiral-invariant masses ($m_0^{(1)} > 600$ MeV and $m_0^{(2)}$ in the range 650--1250 MeV) necessary for successful fits. As demonstrated in Appendix~\ref{sec_su2PDM} for the SU(2) parity doublet case, increasing chiral-invariant masses enhances mixing between chiral partners, which systematically reduces the axial charge---a feature that persists in our SU(3) framework. While this discrepancy appears problematic, it can be resolved by including higher-order derivative terms that contribute to the axial charge without affecting the mass spectrum. These derivative couplings, which we have not included in this leading-order analysis provide additional contribution to the axial charge~\cite{Yamazaki:2018stk,Kummer:2025kch} while leaving the static baryon masses unchanged. In Appendix~\ref{append_3} we also presents the extended Goldberger-Treiman relation connecting the axial charges to pion-nucleon couplings in our $SU(3)$ parity doublet model.

Several important extensions of this work is possible for future investigation. First, the suppressed axial charge $g_A \approx 0.1$ can be remedied by incorporating higher-order derivative couplings, which contribute to the axial current without modifying the mass spectrum~\cite{Yamazaki:2018stk,Kummer:2025kch}. With realistic $g_A$ values in hand, a systematic calculation of meson-baryon coupling constants and decay widths becomes feasible through the extended Goldberger-Treiman relation derived in Appendix~C. This would enable quantitative predictions for resonance widths, particularly for the Roper $N(1440)$ and the hyperon excited states, providing further tests of our model against experimental data. Second, a proper treatment of the $\eta$ meson and $\eta'$ meson requires incorporating $U_A(1)$ anomaly effects, which would allow us to study the $N^*(1535)$-$N$-$\eta$ coupling and address the strong affinity of $N^*(1535)$ for the $\eta N$ channel. Third, the connection between our mirror assignment and the five-quark (pentaquark) structure suggests that explicit pentaquark degrees of freedom might play a role in describing states with significant meson-baryon molecular components. Finally, extending the present framework to finite density and temperature would enable applications to neutron star matter and heavy-ion collisions, where chiral symmetry restoration through parity doubling is expected to play a crucial role.

\appendix
\section{Decomposition of chiral multiplets}\label{append_1}
In this part, we demonstrate that the chiral representations $ (3, \bar{3}), (\bar{3}, 3)$ correspond to $\psi$ field and $(3,6), (6,3)$ correspond to $\eta$ field used in our model naturally emerge from considering the baryon-meson coupling in the context of a five-quark state picture.
We start by considering the tensor product:
\begin{align}
\left[ (3, 1) \oplus (1, 3) \right]^3 \otimes (3, \bar{3})
\end{align}
In this expression, $\left[ (3, 1) \oplus (1, 3) \right]^3$ denote for the baryon formed from three valence quarks and the $(3, \bar{3})$ is from representation of the meson field. First, we decompose the cubic component as
\begin{align}
&\left[ (3, 1) \oplus (1, 3) \right]^3 \nonumber\\
&= (10, 1) \oplus 3(8, 1) \oplus (6, 3) \oplus 2(1,1) \oplus 3(\bar{3}, 3) \nonumber\\
&\quad \oplus 2( 6, 3)  \oplus 3 (3, \bar{3}) \oplus 3(3, 6)  \oplus (1, 8) \oplus (1, 10) 
\end{align}
Next, we compute the tensor products of each component with $(3, \bar{3})$. The results include the following terms
\begin{equation}
\begin{aligned}
(10, 1) \otimes (3, \bar{3}) &=  (15, \bar{3}) \oplus (\bar{6}, \bar{3}),\\
(8, 1) \otimes (3, \bar{3}) &= (15, \bar{3})  \oplus  (\bar{6}, \bar{3}),\\
(3, \bar{3}) \otimes (3, \bar{3}) &=  (6, \bar{6}) \oplus  (6, 3) \oplus (\bar{3}, \bar{6}) \oplus (\bar{3}, 3) ,\\
(1, 10) \otimes (3,\bar{3}) &= (3, 24) \oplus (3, 6),\\
(1, 1) \otimes (3, \bar{3}) &= (3, \bar{3})
\end{aligned}
\end{equation}
In such calculations, we can find the $(3, \bar{3}), (\bar{3}, 3)$ and $(3, 6), (6, 3), (\bar{3}, \bar{6}), (\bar{6}, \bar{3})$ chiral representations are naturally included from the quark structure consideration, validating our choice of fields for describing the octet baryons within the chiral framework.

\section{SU(2) parity doublet} \label{sec_su2PDM}
In the simple $SU(2)$ parity doublet model, the Lagrangian is given as
\begin{align}
\mathcal{L} =& \bar{N}_1 i \slashed{\partial} N_1 - g_1 \bar{N}_1 M N_1   \nonumber\\
 &+ \bar{N}_2 i \slashed{\partial} N_2  - g_2\bar{N}_2 M^{\dagger} N_2  \nonumber\\
 & - m_0 (\bar{N}_1\gamma_5 N_2 - \bar{N}_2 \gamma_5 N_1) + \mathcal{L}_{{\rm meson}}
\end{align}
with the matrix calculated as
\begin{align}
\hat{M}  = 
\left[
\begin{matrix}
g_1\alpha & m_0 \gamma_5 \\
- m_0 \gamma_5 & g_2 \alpha \\
\end{matrix}\right]
\end{align}
This mass matrix can be diagonalized  with the mixing angle
\begin{align}
{\rm tan} 2\theta = \frac{2 m_0}{ (g_1 + g_2) \alpha}.
\end{align}
and the corresponding masses of two states are given as
\begin{align}
m_\pm = \frac{1}{2} \left( \sqrt{ (g_1 + g_2)^2\alpha^2 + 4m_0^2 }  \pm (g_1 - g_2)\alpha \right)
\end{align}
We can rewrite the mass formula to obtain $g_1, g_2$ values from the vacuum inputs with $m_+ = 939$ MeV and $m_-= 1535$ MeV
\begin{equation}
g_{1,2}=\frac{1}{2 f_\pi}\left(\sqrt{\left(m_{-}+m_{+}\right)^2-4 m_0^2} \pm\left(m_{-}-m_{+}\right)\right)
\end{equation}
We also obtain its axial charges as
\begin{align}
g_A = \left[
\begin{matrix}
{\rm cos} 2\theta & - {\rm sin} 2\theta \gamma_5 \\
- {\rm sin} 2\theta \gamma_5 & -{\rm cos} 2\theta \\
\end{matrix}\right]
\end{align}
with (1, 1) component $g_A^{++} = {\rm cos} 2\theta$ denote for the axial charge for the ground state nucleon.  Also, we can see that the axial charge from the negative parity nucleon is always the minus of that of the positive parity nucleon. Since there is a mixing between two nucleons, the axial charge is always smaller than unity. Besides, as the increasing of the $m_0$ values, the mixing angle become larger and the corresponding axial charge of nucleon will decrease.
\begin{figure}[htb]
\centering
\includegraphics[width=0.9\hsize]{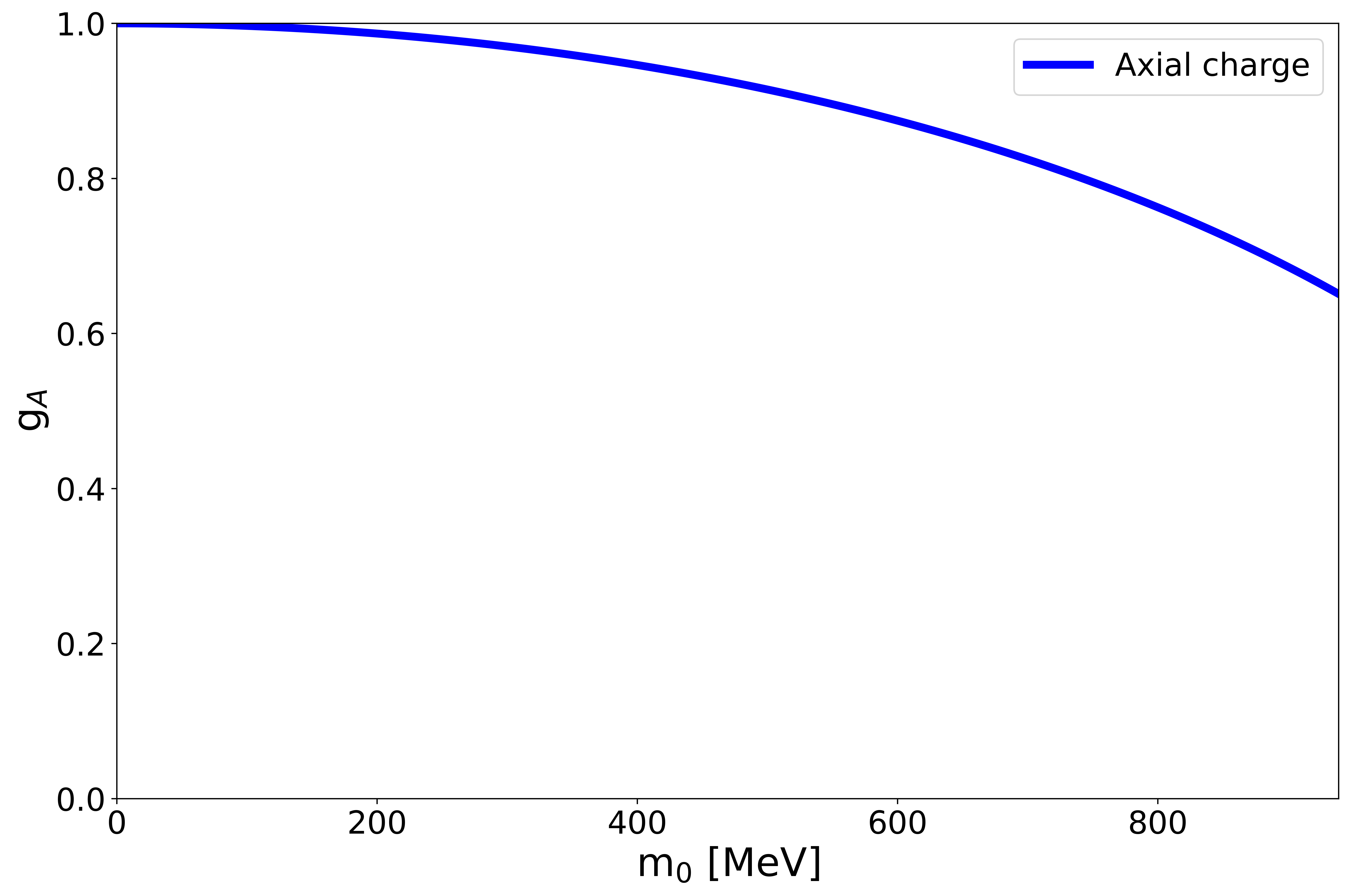}
\caption{Ground state nucleon axial charge in $SU(2)$ parity doublet model.}
\label{fig:axial_SU2}
\end{figure}
\begin{table} [tb]
\centering
\caption{Values of coupling constants $g_1, g_2$, mixing angle $\theta$ and corresponding ground state nucleon axial charge $g_A$ for different choices of chiral invariant mass $m_0$.}
\begin{tabular}{c|ccccc}
\hline \hline$m_0[\mathrm{MeV}]$ & 200 & 400 & 600 & 800 \\
\hline
$g_1$ & 16.33 & 15.79 & 14.84 & 13.35  \\
$g_2$ & 9.92 & 9.38 & 8.43 & 6.94  \\
$\theta$ & 4.7° & 9.4° & 14.5° & 20.1° \\
$g_A$ & 0.98 & 0.94 & 0.87 & 0.76\\
\hline \hline
\end{tabular}
\end{table}

\section{Extended Goldberger-Treiman relation}\label{append_3}
It is useful to also see the axial charges which are related to the one pion Yukawa couplings through the Goldberger-Treiman relation. The axial coupling are determined by the chiral representation from the kinetic terms 
which is calculated in the matrix form as
\begin{align}
g_A &=U^T M_A  U \\
M_A &= \left[\begin{matrix}
-1  & 
0 & 0 & 0\\
&   \frac{5}{3} & 0&0\\
&   &1 & 0\\
& & & -\frac{5}{3}
\end{matrix} \right] 
\end{align}
where $U$ denotes the eigenvector corresponding to $N(939)$. Our calculation yields $g_A \approx 0.1$, smaller than the experimental value of $g_A^{\text{exp}} = 1.27$. 
This suppression of the axial charge can be understood from the large chiral-invariant masses required in our model. As shown in Fig.~\ref{fig-groundratio} and Fig.~\ref{fig-mix_N}, successful fits to the baryon spectrum require both $m_0^{(1)}$ and $m_0^{(2)}$ to be large. These large chiral-invariant masses generate substantial mixing between the original fields and their mirror partners, analogous to the case in the $SU(2)$ parity doublet model. As demonstrated in Fig.~\ref{fig:axial_SU2} for the $SU(2)$ case, increasing the mixing angle systematically reduces the axial charge—a feature that persists in our $SU(3)$ framework.

This discrepancy between theoretical and experimental axial charges is not a fundamental limitation of the model. Higher-order derivative terms, which we have not included in this analysis, can contribute to the axial charge without affecting the mass spectrum. These derivative couplings contribute additional momentum-dependent terms to the axial current while leaving the static baryon masses unchanged, providing a natural mechanism to reconcile the model with experimental data in a more complete treatment.

We also obtain the interaction term for nucleons which is calculated from the Lagrangian in the form $\bar{N} C_{\pi NN} i \gamma_5 \pi N$, with $C_{\pi NN}$ is 
\begin{align}
C_{\pi NN} = \left[\begin{matrix}
-g_1  & -\frac{1}{\sqrt{6}}y_1 & 0 & 0\\
        &   \frac{5}{6}g_3         & 0 &0\\
        &                               &-g_2 & - \frac{1}{\sqrt{6}}y_2 \\
        &                               &  & -\frac{5}{6}g_4
\end{matrix} \right] .
\end{align}
By using  matrix $M_A$, the mass matrix of nucleon $\hat{M}_N$ and the interaction terms, we find the extended Goldberger-Treiman relation
\begin{align}
C_{\pi NN} = \frac{1}{2f_{\pi}} \left\{ M_A, \hat{M}_N\right\}.
\end{align}


\bibliography{ref_3fPDM_2022.bib}

\begin{thebibliography}{71}%
\makeatletter
\providecommand \@ifxundefined [1]{%
 \@ifx{#1\undefined}
}%
\providecommand \@ifnum [1]{%
 \ifnum #1\expandafter \@firstoftwo
 \else \expandafter \@secondoftwo
 \fi
}%
\providecommand \@ifx [1]{%
 \ifx #1\expandafter \@firstoftwo
 \else \expandafter \@secondoftwo
 \fi
}%
\providecommand \natexlab [1]{#1}%
\providecommand \enquote  [1]{``#1''}%
\providecommand \bibnamefont  [1]{#1}%
\providecommand \bibfnamefont [1]{#1}%
\providecommand \citenamefont [1]{#1}%
\providecommand \href@noop [0]{\@secondoftwo}%
\providecommand \href [0]{\begingroup \@sanitize@url \@href}%
\providecommand \@href[1]{\@@startlink{#1}\@@href}%
\providecommand \@@href[1]{\endgroup#1\@@endlink}%
\providecommand \@sanitize@url [0]{\catcode `\\12\catcode `\$12\catcode
  `\&12\catcode `\#12\catcode `\^12\catcode `\_12\catcode `\%12\relax}%
\providecommand \@@startlink[1]{}%
\providecommand \@@endlink[0]{}%
\providecommand \url  [0]{\begingroup\@sanitize@url \@url }%
\providecommand \@url [1]{\endgroup\@href {#1}{\urlprefix }}%
\providecommand \urlprefix  [0]{URL }%
\providecommand \Eprint [0]{\href }%
\providecommand \doibase [0]{https://doi.org/}%
\providecommand \selectlanguage [0]{\@gobble}%
\providecommand \bibinfo  [0]{\@secondoftwo}%
\providecommand \bibfield  [0]{\@secondoftwo}%
\providecommand \translation [1]{[#1]}%
\providecommand \BibitemOpen [0]{}%
\providecommand \bibitemStop [0]{}%
\providecommand \bibitemNoStop [0]{.\EOS\space}%
\providecommand \EOS [0]{\spacefactor3000\relax}%
\providecommand \BibitemShut  [1]{\csname bibitem#1\endcsname}%
\let\auto@bib@innerbib\@empty
\bibitem [{\citenamefont {Detar}\ and\ \citenamefont
  {Kunihiro}(1989)}]{Detar:1988kn}%
  \BibitemOpen
  \bibfield  {author} {\bibinfo {author} {\bibfnamefont {C.~E.}\ \bibnamefont
  {Detar}}\ and\ \bibinfo {author} {\bibfnamefont {T.}~\bibnamefont
  {Kunihiro}},\ }\bibfield  {title} {\bibinfo {title} {{Linear $\sigma$ Model
  With Parity Doubling}},\ }\href {https://doi.org/10.1103/PhysRevD.39.2805}
  {\bibfield  {journal} {\bibinfo  {journal} {Phys. Rev. D}\ }\textbf {\bibinfo
  {volume} {39}},\ \bibinfo {pages} {2805} (\bibinfo {year}
  {1989})}\BibitemShut {NoStop}%
\bibitem [{\citenamefont {Jido}\ \emph
  {et~al.}(2000{\natexlab{a}})\citenamefont {Jido}, \citenamefont {Nemoto},
  \citenamefont {Oka},\ and\ \citenamefont {Hosaka}}]{Jido:1998av}%
  \BibitemOpen
  \bibfield  {author} {\bibinfo {author} {\bibfnamefont {D.}~\bibnamefont
  {Jido}}, \bibinfo {author} {\bibfnamefont {Y.}~\bibnamefont {Nemoto}},
  \bibinfo {author} {\bibfnamefont {M.}~\bibnamefont {Oka}},\ and\ \bibinfo
  {author} {\bibfnamefont {A.}~\bibnamefont {Hosaka}},\ }\bibfield  {title}
  {\bibinfo {title} {{Chiral symmetry for positive and negative parity
  nucleons}},\ }\href {https://doi.org/10.1016/S0375-9474(99)00844-1}
  {\bibfield  {journal} {\bibinfo  {journal} {Nucl. Phys. A}\ }\textbf
  {\bibinfo {volume} {671}},\ \bibinfo {pages} {471} (\bibinfo {year}
  {2000}{\natexlab{a}})},\ \Eprint {https://arxiv.org/abs/hep-ph/9805306}
  {arXiv:hep-ph/9805306} \BibitemShut {NoStop}%
\bibitem [{\citenamefont {Jido}\ \emph {et~al.}(2001)\citenamefont {Jido},
  \citenamefont {Oka},\ and\ \citenamefont {Hosaka}}]{Jido:2001nt}%
  \BibitemOpen
  \bibfield  {author} {\bibinfo {author} {\bibfnamefont {D.}~\bibnamefont
  {Jido}}, \bibinfo {author} {\bibfnamefont {M.}~\bibnamefont {Oka}},\ and\
  \bibinfo {author} {\bibfnamefont {A.}~\bibnamefont {Hosaka}},\ }\bibfield
  {title} {\bibinfo {title} {{Chiral symmetry of baryons}},\ }\href
  {https://doi.org/10.1143/PTP.106.873} {\bibfield  {journal} {\bibinfo
  {journal} {Prog. Theor. Phys.}\ }\textbf {\bibinfo {volume} {106}},\ \bibinfo
  {pages} {873} (\bibinfo {year} {2001})},\ \Eprint
  {https://arxiv.org/abs/hep-ph/0110005} {arXiv:hep-ph/0110005} \BibitemShut
  {NoStop}%
\bibitem [{\citenamefont {Nagata}\ \emph {et~al.}(2008)\citenamefont {Nagata},
  \citenamefont {Hosaka},\ and\ \citenamefont {Dmitrasinovic}}]{Nagata:2008xf}%
  \BibitemOpen
  \bibfield  {author} {\bibinfo {author} {\bibfnamefont {K.}~\bibnamefont
  {Nagata}}, \bibinfo {author} {\bibfnamefont {A.}~\bibnamefont {Hosaka}},\
  and\ \bibinfo {author} {\bibfnamefont {V.}~\bibnamefont {Dmitrasinovic}},\
  }\bibfield  {title} {\bibinfo {title} {{pi N and pi pi N Couplings of the
  Delta(1232) and its Chiral Partners}},\ }\href
  {https://doi.org/10.1103/PhysRevLett.101.092001} {\bibfield  {journal}
  {\bibinfo  {journal} {Phys. Rev. Lett.}\ }\textbf {\bibinfo {volume} {101}},\
  \bibinfo {pages} {092001} (\bibinfo {year} {2008})},\ \Eprint
  {https://arxiv.org/abs/0804.3185} {arXiv:0804.3185 [hep-ph]} \BibitemShut
  {NoStop}%
\bibitem [{\citenamefont {Sasaki}\ and\ \citenamefont
  {Mishustin}(2010)}]{Sasaki:2010bp}%
  \BibitemOpen
  \bibfield  {author} {\bibinfo {author} {\bibfnamefont {C.}~\bibnamefont
  {Sasaki}}\ and\ \bibinfo {author} {\bibfnamefont {I.}~\bibnamefont
  {Mishustin}},\ }\bibfield  {title} {\bibinfo {title} {{Thermodynamics of
  dense hadronic matter in a parity doublet model}},\ }\href
  {https://doi.org/10.1103/PhysRevC.82.035204} {\bibfield  {journal} {\bibinfo
  {journal} {Phys. Rev. C}\ }\textbf {\bibinfo {volume} {82}},\ \bibinfo
  {pages} {035204} (\bibinfo {year} {2010})},\ \Eprint
  {https://arxiv.org/abs/1005.4811} {arXiv:1005.4811 [hep-ph]} \BibitemShut
  {NoStop}%
\bibitem [{\citenamefont {Gallas}\ and\ \citenamefont
  {Giacosa}(2014)}]{Gallas:2013ipa}%
  \BibitemOpen
  \bibfield  {author} {\bibinfo {author} {\bibfnamefont {S.}~\bibnamefont
  {Gallas}}\ and\ \bibinfo {author} {\bibfnamefont {F.}~\bibnamefont
  {Giacosa}},\ }\bibfield  {title} {\bibinfo {title} {{Mirror versus naive
  assignment in chiral models for the nucleon}},\ }\href
  {https://doi.org/10.1142/S0217751X14500985} {\bibfield  {journal} {\bibinfo
  {journal} {Int. J. Mod. Phys. A}\ }\textbf {\bibinfo {volume} {29}},\
  \bibinfo {pages} {1450098} (\bibinfo {year} {2014})},\ \Eprint
  {https://arxiv.org/abs/1308.4817} {arXiv:1308.4817 [hep-ph]} \BibitemShut
  {NoStop}%
\bibitem [{\citenamefont {Aarts}\ \emph {et~al.}(2015)\citenamefont {Aarts},
  \citenamefont {Allton}, \citenamefont {Hands}, \citenamefont {J\"ager},
  \citenamefont {Praki},\ and\ \citenamefont {Skullerud}}]{Aarts:2015mma}%
  \BibitemOpen
  \bibfield  {author} {\bibinfo {author} {\bibfnamefont {G.}~\bibnamefont
  {Aarts}}, \bibinfo {author} {\bibfnamefont {C.}~\bibnamefont {Allton}},
  \bibinfo {author} {\bibfnamefont {S.}~\bibnamefont {Hands}}, \bibinfo
  {author} {\bibfnamefont {B.}~\bibnamefont {J\"ager}}, \bibinfo {author}
  {\bibfnamefont {C.}~\bibnamefont {Praki}},\ and\ \bibinfo {author}
  {\bibfnamefont {J.-I.}\ \bibnamefont {Skullerud}},\ }\bibfield  {title}
  {\bibinfo {title} {{Nucleons and parity doubling across the deconfinement
  transition}},\ }\href {https://doi.org/10.1103/PhysRevD.92.014503} {\bibfield
   {journal} {\bibinfo  {journal} {Phys. Rev. D}\ }\textbf {\bibinfo {volume}
  {92}},\ \bibinfo {pages} {014503} (\bibinfo {year} {2015})},\ \Eprint
  {https://arxiv.org/abs/1502.03603} {arXiv:1502.03603 [hep-lat]} \BibitemShut
  {NoStop}%
\bibitem [{\citenamefont {Aarts}\ \emph {et~al.}(2017)\citenamefont {Aarts},
  \citenamefont {Allton}, \citenamefont {De~Boni}, \citenamefont {Hands},
  \citenamefont {J\"ager}, \citenamefont {Praki},\ and\ \citenamefont
  {Skullerud}}]{Aarts:2017rrl}%
  \BibitemOpen
  \bibfield  {author} {\bibinfo {author} {\bibfnamefont {G.}~\bibnamefont
  {Aarts}}, \bibinfo {author} {\bibfnamefont {C.}~\bibnamefont {Allton}},
  \bibinfo {author} {\bibfnamefont {D.}~\bibnamefont {De~Boni}}, \bibinfo
  {author} {\bibfnamefont {S.}~\bibnamefont {Hands}}, \bibinfo {author}
  {\bibfnamefont {B.}~\bibnamefont {J\"ager}}, \bibinfo {author} {\bibfnamefont
  {C.}~\bibnamefont {Praki}},\ and\ \bibinfo {author} {\bibfnamefont {J.-I.}\
  \bibnamefont {Skullerud}},\ }\bibfield  {title} {\bibinfo {title} {{Light
  baryons below and above the deconfinement transition: medium effects and
  parity doubling}},\ }\href {https://doi.org/10.1007/JHEP06(2017)034}
  {\bibfield  {journal} {\bibinfo  {journal} {JHEP}\ }\textbf {\bibinfo
  {volume} {06}},\ \bibinfo {pages} {034}},\ \Eprint
  {https://arxiv.org/abs/1703.09246} {arXiv:1703.09246 [hep-lat]} \BibitemShut
  {NoStop}%
\bibitem [{\citenamefont {Aarts}\ \emph {et~al.}(2019)\citenamefont {Aarts},
  \citenamefont {Allton}, \citenamefont {De~Boni},\ and\ \citenamefont
  {J\"ager}}]{Aarts:2018glk}%
  \BibitemOpen
  \bibfield  {author} {\bibinfo {author} {\bibfnamefont {G.}~\bibnamefont
  {Aarts}}, \bibinfo {author} {\bibfnamefont {C.}~\bibnamefont {Allton}},
  \bibinfo {author} {\bibfnamefont {D.}~\bibnamefont {De~Boni}},\ and\ \bibinfo
  {author} {\bibfnamefont {B.}~\bibnamefont {J\"ager}},\ }\bibfield  {title}
  {\bibinfo {title} {{Hyperons in thermal QCD: A lattice view}},\ }\href
  {https://doi.org/10.1103/PhysRevD.99.074503} {\bibfield  {journal} {\bibinfo
  {journal} {Phys. Rev. D}\ }\textbf {\bibinfo {volume} {99}},\ \bibinfo
  {pages} {074503} (\bibinfo {year} {2019})},\ \Eprint
  {https://arxiv.org/abs/1812.07393} {arXiv:1812.07393 [hep-lat]} \BibitemShut
  {NoStop}%
\bibitem [{\citenamefont {Takahashi}\ and\ \citenamefont
  {Kunihiro}(2008)}]{Takahashi:2008fy}%
  \BibitemOpen
  \bibfield  {author} {\bibinfo {author} {\bibfnamefont {T.~T.}\ \bibnamefont
  {Takahashi}}\ and\ \bibinfo {author} {\bibfnamefont {T.}~\bibnamefont
  {Kunihiro}},\ }\bibfield  {title} {\bibinfo {title} {{Axial charges of
  N(1535) and N(1650) in lattice QCD with two flavors of dynamical quarks}},\
  }\href {https://doi.org/10.1103/PhysRevD.78.011503} {\bibfield  {journal}
  {\bibinfo  {journal} {Phys. Rev. D}\ }\textbf {\bibinfo {volume} {78}},\
  \bibinfo {pages} {011503} (\bibinfo {year} {2008})},\ \Eprint
  {https://arxiv.org/abs/0801.4707} {arXiv:0801.4707 [hep-lat]} \BibitemShut
  {NoStop}%
\bibitem [{\citenamefont {Weinberg}(1968)}]{Weinberg:1968de}%
  \BibitemOpen
  \bibfield  {author} {\bibinfo {author} {\bibfnamefont {S.}~\bibnamefont
  {Weinberg}},\ }\bibfield  {title} {\bibinfo {title} {{Nonlinear realizations
  of chiral symmetry}},\ }\href {https://doi.org/10.1103/PhysRev.166.1568}
  {\bibfield  {journal} {\bibinfo  {journal} {Phys. Rev.}\ }\textbf {\bibinfo
  {volume} {166}},\ \bibinfo {pages} {1568} (\bibinfo {year}
  {1968})}\BibitemShut {NoStop}%
\bibitem [{\citenamefont {Bando}\ \emph {et~al.}(1988)\citenamefont {Bando},
  \citenamefont {Kugo},\ and\ \citenamefont {Yamawaki}}]{Bando:1987br}%
  \BibitemOpen
  \bibfield  {author} {\bibinfo {author} {\bibfnamefont {M.}~\bibnamefont
  {Bando}}, \bibinfo {author} {\bibfnamefont {T.}~\bibnamefont {Kugo}},\ and\
  \bibinfo {author} {\bibfnamefont {K.}~\bibnamefont {Yamawaki}},\ }\bibfield
  {title} {\bibinfo {title} {{Nonlinear Realization and Hidden Local
  Symmetries}},\ }\href {https://doi.org/10.1016/0370-1573(88)90019-1}
  {\bibfield  {journal} {\bibinfo  {journal} {Phys. Rept.}\ }\textbf {\bibinfo
  {volume} {164}},\ \bibinfo {pages} {217} (\bibinfo {year}
  {1988})}\BibitemShut {NoStop}%
\bibitem [{\citenamefont {Buscher}(1988)}]{Buscher:1987qj}%
  \BibitemOpen
  \bibfield  {author} {\bibinfo {author} {\bibfnamefont {T.~H.}\ \bibnamefont
  {Buscher}},\ }\bibfield  {title} {\bibinfo {title} {{Path Integral Derivation
  of Quantum Duality in Nonlinear Sigma Models}},\ }\href
  {https://doi.org/10.1016/0370-2693(88)90602-8} {\bibfield  {journal}
  {\bibinfo  {journal} {Phys. Lett. B}\ }\textbf {\bibinfo {volume} {201}},\
  \bibinfo {pages} {466} (\bibinfo {year} {1988})}\BibitemShut {NoStop}%
\bibitem [{\citenamefont {Gasser}\ and\ \citenamefont
  {Leutwyler}(1988)}]{Gasser:1987zq}%
  \BibitemOpen
  \bibfield  {author} {\bibinfo {author} {\bibfnamefont {J.}~\bibnamefont
  {Gasser}}\ and\ \bibinfo {author} {\bibfnamefont {H.}~\bibnamefont
  {Leutwyler}},\ }\bibfield  {title} {\bibinfo {title} {{Spontaneously broken
  symmetries: Effective lagrangians at finite volume}},\ }\href
  {https://doi.org/10.1016/0550-3213(88)90107-1} {\bibfield  {journal}
  {\bibinfo  {journal} {Nucl. Phys. B}\ }\textbf {\bibinfo {volume} {307}},\
  \bibinfo {pages} {763} (\bibinfo {year} {1988})}\BibitemShut {NoStop}%
\bibitem [{\citenamefont {Coleman}\ \emph {et~al.}(1969)\citenamefont
  {Coleman}, \citenamefont {Wess},\ and\ \citenamefont
  {Zumino}}]{Coleman:1969sm}%
  \BibitemOpen
  \bibfield  {author} {\bibinfo {author} {\bibfnamefont {S.~R.}\ \bibnamefont
  {Coleman}}, \bibinfo {author} {\bibfnamefont {J.}~\bibnamefont {Wess}},\ and\
  \bibinfo {author} {\bibfnamefont {B.}~\bibnamefont {Zumino}},\ }\bibfield
  {title} {\bibinfo {title} {{Structure of phenomenological Lagrangians. 1.}},\
  }\href {https://doi.org/10.1103/PhysRev.177.2239} {\bibfield  {journal}
  {\bibinfo  {journal} {Phys. Rev.}\ }\textbf {\bibinfo {volume} {177}},\
  \bibinfo {pages} {2239} (\bibinfo {year} {1969})}\BibitemShut {NoStop}%
\bibitem [{\citenamefont {Callan}\ \emph {et~al.}(1969)\citenamefont {Callan},
  \citenamefont {Coleman}, \citenamefont {Wess},\ and\ \citenamefont
  {Zumino}}]{Callan:1969sn}%
  \BibitemOpen
  \bibfield  {author} {\bibinfo {author} {\bibfnamefont {C.~G.}\ \bibnamefont
  {Callan}, \bibfnamefont {Jr.}}, \bibinfo {author} {\bibfnamefont {S.~R.}\
  \bibnamefont {Coleman}}, \bibinfo {author} {\bibfnamefont {J.}~\bibnamefont
  {Wess}},\ and\ \bibinfo {author} {\bibfnamefont {B.}~\bibnamefont {Zumino}},\
  }\bibfield  {title} {\bibinfo {title} {{Structure of phenomenological
  Lagrangians. 2.}},\ }\href {https://doi.org/10.1103/PhysRev.177.2247}
  {\bibfield  {journal} {\bibinfo  {journal} {Phys. Rev.}\ }\textbf {\bibinfo
  {volume} {177}},\ \bibinfo {pages} {2247} (\bibinfo {year}
  {1969})}\BibitemShut {NoStop}%
\bibitem [{\citenamefont {Papazoglou}\ \emph {et~al.}(1999)\citenamefont
  {Papazoglou}, \citenamefont {Zschiesche}, \citenamefont {Schramm},
  \citenamefont {Schaffner-Bielich}, \citenamefont {Stoecker},\ and\
  \citenamefont {Greiner}}]{Papazoglou:1998vr}%
  \BibitemOpen
  \bibfield  {author} {\bibinfo {author} {\bibfnamefont {P.}~\bibnamefont
  {Papazoglou}}, \bibinfo {author} {\bibfnamefont {D.}~\bibnamefont
  {Zschiesche}}, \bibinfo {author} {\bibfnamefont {S.}~\bibnamefont {Schramm}},
  \bibinfo {author} {\bibfnamefont {J.}~\bibnamefont {Schaffner-Bielich}},
  \bibinfo {author} {\bibfnamefont {H.}~\bibnamefont {Stoecker}},\ and\
  \bibinfo {author} {\bibfnamefont {W.}~\bibnamefont {Greiner}},\ }\bibfield
  {title} {\bibinfo {title} {{Nuclei in a chiral SU(3) model}},\ }\href
  {https://doi.org/10.1103/PhysRevC.59.411} {\bibfield  {journal} {\bibinfo
  {journal} {Phys. Rev. C}\ }\textbf {\bibinfo {volume} {59}},\ \bibinfo
  {pages} {411} (\bibinfo {year} {1999})},\ \Eprint
  {https://arxiv.org/abs/nucl-th/9806087} {arXiv:nucl-th/9806087} \BibitemShut
  {NoStop}%
\bibitem [{\citenamefont {Mishra}\ \emph {et~al.}(2004)\citenamefont {Mishra},
  \citenamefont {Balazs}, \citenamefont {Zschiesche}, \citenamefont {Schramm},
  \citenamefont {Stoecker},\ and\ \citenamefont {Greiner}}]{Mishra:2003tr}%
  \BibitemOpen
  \bibfield  {author} {\bibinfo {author} {\bibfnamefont {A.}~\bibnamefont
  {Mishra}}, \bibinfo {author} {\bibfnamefont {K.}~\bibnamefont {Balazs}},
  \bibinfo {author} {\bibfnamefont {D.}~\bibnamefont {Zschiesche}}, \bibinfo
  {author} {\bibfnamefont {S.}~\bibnamefont {Schramm}}, \bibinfo {author}
  {\bibfnamefont {H.}~\bibnamefont {Stoecker}},\ and\ \bibinfo {author}
  {\bibfnamefont {W.}~\bibnamefont {Greiner}},\ }\bibfield  {title} {\bibinfo
  {title} {{Effects of Dirac sea polarization on hadronic properties: A Chiral
  SU(3) approach}},\ }\href {https://doi.org/10.1103/PhysRevC.69.024903}
  {\bibfield  {journal} {\bibinfo  {journal} {Phys. Rev. C}\ }\textbf {\bibinfo
  {volume} {69}},\ \bibinfo {pages} {024903} (\bibinfo {year} {2004})},\
  \Eprint {https://arxiv.org/abs/nucl-th/0308064} {arXiv:nucl-th/0308064}
  \BibitemShut {NoStop}%
\bibitem [{\citenamefont {Steinheimer}\ \emph {et~al.}(2011)\citenamefont
  {Steinheimer}, \citenamefont {Schramm},\ and\ \citenamefont
  {Stocker}}]{Steinheimer:2011ea}%
  \BibitemOpen
  \bibfield  {author} {\bibinfo {author} {\bibfnamefont {J.}~\bibnamefont
  {Steinheimer}}, \bibinfo {author} {\bibfnamefont {S.}~\bibnamefont
  {Schramm}},\ and\ \bibinfo {author} {\bibfnamefont {H.}~\bibnamefont
  {Stocker}},\ }\bibfield  {title} {\bibinfo {title} {{The hadronic SU(3)
  Parity Doublet Model for Dense Matter, its extension to quarks and the
  strange equation of state}},\ }\href
  {https://doi.org/10.1103/PhysRevC.84.045208} {\bibfield  {journal} {\bibinfo
  {journal} {Phys. Rev. C}\ }\textbf {\bibinfo {volume} {84}},\ \bibinfo
  {pages} {045208} (\bibinfo {year} {2011})},\ \Eprint
  {https://arxiv.org/abs/1108.2596} {arXiv:1108.2596 [hep-ph]} \BibitemShut
  {NoStop}%
\bibitem [{\citenamefont {Dexheimer}\ \emph {et~al.}(2013)\citenamefont
  {Dexheimer}, \citenamefont {Steinheimer}, \citenamefont {Negreiros},\ and\
  \citenamefont {Schramm}}]{Dexheimer:2012eu}%
  \BibitemOpen
  \bibfield  {author} {\bibinfo {author} {\bibfnamefont {V.}~\bibnamefont
  {Dexheimer}}, \bibinfo {author} {\bibfnamefont {J.}~\bibnamefont
  {Steinheimer}}, \bibinfo {author} {\bibfnamefont {R.}~\bibnamefont
  {Negreiros}},\ and\ \bibinfo {author} {\bibfnamefont {S.}~\bibnamefont
  {Schramm}},\ }\bibfield  {title} {\bibinfo {title} {{Hybrid Stars in an SU(3)
  parity doublet model}},\ }\href {https://doi.org/10.1103/PhysRevC.87.015804}
  {\bibfield  {journal} {\bibinfo  {journal} {Phys. Rev. C}\ }\textbf {\bibinfo
  {volume} {87}},\ \bibinfo {pages} {015804} (\bibinfo {year} {2013})},\
  \Eprint {https://arxiv.org/abs/1206.3086} {arXiv:1206.3086 [astro-ph.HE]}
  \BibitemShut {NoStop}%
\bibitem [{\citenamefont {Fraga}\ \emph {et~al.}(2023)\citenamefont {Fraga},
  \citenamefont {da~Mata},\ and\ \citenamefont
  {Schaffner-Bielich}}]{Fraga:2023wtd}%
  \BibitemOpen
  \bibfield  {author} {\bibinfo {author} {\bibfnamefont {E.~S.}\ \bibnamefont
  {Fraga}}, \bibinfo {author} {\bibfnamefont {R.}~\bibnamefont {da~Mata}},\
  and\ \bibinfo {author} {\bibfnamefont {J.}~\bibnamefont
  {Schaffner-Bielich}},\ }\bibfield  {title} {\bibinfo {title} {{SU(3) parity
  doubling in cold neutron star matter}},\ }\href
  {https://doi.org/10.1103/PhysRevD.108.116003} {\bibfield  {journal} {\bibinfo
   {journal} {Phys. Rev. D}\ }\textbf {\bibinfo {volume} {108}},\ \bibinfo
  {pages} {116003} (\bibinfo {year} {2023})},\ \Eprint
  {https://arxiv.org/abs/2309.02368} {arXiv:2309.02368 [hep-ph]} \BibitemShut
  {NoStop}%
\bibitem [{\citenamefont {Hatsuda}\ and\ \citenamefont
  {Prakash}(1989)}]{Hatsuda:1988mv}%
  \BibitemOpen
  \bibfield  {author} {\bibinfo {author} {\bibfnamefont {T.}~\bibnamefont
  {Hatsuda}}\ and\ \bibinfo {author} {\bibfnamefont {M.}~\bibnamefont
  {Prakash}},\ }\bibfield  {title} {\bibinfo {title} {{Parity Doubling of the
  Nucleon and First Order Chiral Transition in Dense Matter}},\ }\href
  {https://doi.org/10.1016/0370-2693(89)91040-X} {\bibfield  {journal}
  {\bibinfo  {journal} {Phys. Lett. B}\ }\textbf {\bibinfo {volume} {224}},\
  \bibinfo {pages} {11} (\bibinfo {year} {1989})}\BibitemShut {NoStop}%
\bibitem [{\citenamefont {Manohar}\ and\ \citenamefont
  {Georgi}(1984)}]{Manohar:1983md}%
  \BibitemOpen
  \bibfield  {author} {\bibinfo {author} {\bibfnamefont {A.}~\bibnamefont
  {Manohar}}\ and\ \bibinfo {author} {\bibfnamefont {H.}~\bibnamefont
  {Georgi}},\ }\bibfield  {title} {\bibinfo {title} {{Chiral Quarks and the
  Nonrelativistic Quark Model}},\ }\href
  {https://doi.org/10.1016/0550-3213(84)90231-1} {\bibfield  {journal}
  {\bibinfo  {journal} {Nucl. Phys. B}\ }\textbf {\bibinfo {volume} {234}},\
  \bibinfo {pages} {189} (\bibinfo {year} {1984})}\BibitemShut {NoStop}%
\bibitem [{\citenamefont {Zschiesche}\ \emph {et~al.}(2007)\citenamefont
  {Zschiesche}, \citenamefont {Tolos}, \citenamefont {Schaffner-Bielich},\ and\
  \citenamefont {Pisarski}}]{Zschiesche:2006zj}%
  \BibitemOpen
  \bibfield  {author} {\bibinfo {author} {\bibfnamefont {D.}~\bibnamefont
  {Zschiesche}}, \bibinfo {author} {\bibfnamefont {L.}~\bibnamefont {Tolos}},
  \bibinfo {author} {\bibfnamefont {J.}~\bibnamefont {Schaffner-Bielich}},\
  and\ \bibinfo {author} {\bibfnamefont {R.~D.}\ \bibnamefont {Pisarski}},\
  }\bibfield  {title} {\bibinfo {title} {{Cold, dense nuclear matter in a SU(2)
  parity doublet model}},\ }\href {https://doi.org/10.1103/PhysRevC.75.055202}
  {\bibfield  {journal} {\bibinfo  {journal} {Phys. Rev. C}\ }\textbf {\bibinfo
  {volume} {75}},\ \bibinfo {pages} {055202} (\bibinfo {year} {2007})},\
  \Eprint {https://arxiv.org/abs/nucl-th/0608044} {arXiv:nucl-th/0608044}
  \BibitemShut {NoStop}%
\bibitem [{\citenamefont {Dexheimer}\ \emph {et~al.}(2008)\citenamefont
  {Dexheimer}, \citenamefont {Schramm},\ and\ \citenamefont
  {Zschiesche}}]{Dexheimer:2007tn}%
  \BibitemOpen
  \bibfield  {author} {\bibinfo {author} {\bibfnamefont {V.}~\bibnamefont
  {Dexheimer}}, \bibinfo {author} {\bibfnamefont {S.}~\bibnamefont {Schramm}},\
  and\ \bibinfo {author} {\bibfnamefont {D.}~\bibnamefont {Zschiesche}},\
  }\bibfield  {title} {\bibinfo {title} {{Nuclear matter and neutron stars in a
  parity doublet model}},\ }\href {https://doi.org/10.1103/PhysRevC.77.025803}
  {\bibfield  {journal} {\bibinfo  {journal} {Phys. Rev. C}\ }\textbf {\bibinfo
  {volume} {77}},\ \bibinfo {pages} {025803} (\bibinfo {year} {2008})},\
  \Eprint {https://arxiv.org/abs/0710.4192} {arXiv:0710.4192 [nucl-th]}
  \BibitemShut {NoStop}%
\bibitem [{\citenamefont {Chen}\ \emph {et~al.}(2010)\citenamefont {Chen},
  \citenamefont {Dmitrasinovic},\ and\ \citenamefont {Hosaka}}]{Chen:2009sf}%
  \BibitemOpen
  \bibfield  {author} {\bibinfo {author} {\bibfnamefont {H.-X.}\ \bibnamefont
  {Chen}}, \bibinfo {author} {\bibfnamefont {V.}~\bibnamefont
  {Dmitrasinovic}},\ and\ \bibinfo {author} {\bibfnamefont {A.}~\bibnamefont
  {Hosaka}},\ }\bibfield  {title} {\bibinfo {title} {{Baryon fields with
  U(L)(3) X U(R)(3) chiral symmetry II: Axial currents of nucleons and
  hyperons}},\ }\href {https://doi.org/10.1103/PhysRevD.81.054002} {\bibfield
  {journal} {\bibinfo  {journal} {Phys. Rev. D}\ }\textbf {\bibinfo {volume}
  {81}},\ \bibinfo {pages} {054002} (\bibinfo {year} {2010})},\ \Eprint
  {https://arxiv.org/abs/0912.4338} {arXiv:0912.4338 [hep-ph]} \BibitemShut
  {NoStop}%
\bibitem [{\citenamefont {Chen}\ \emph {et~al.}(2011)\citenamefont {Chen},
  \citenamefont {Dmitrasinovic},\ and\ \citenamefont {Hosaka}}]{Chen:2010ba}%
  \BibitemOpen
  \bibfield  {author} {\bibinfo {author} {\bibfnamefont {H.-X.}\ \bibnamefont
  {Chen}}, \bibinfo {author} {\bibfnamefont {V.}~\bibnamefont
  {Dmitrasinovic}},\ and\ \bibinfo {author} {\bibfnamefont {A.}~\bibnamefont
  {Hosaka}},\ }\bibfield  {title} {\bibinfo {title} {{Baryon Fields with
  $U_L(3) times U_R(3)$ Chiral Symmetry III: Interactions with Chiral
  $(3,\bar{3})+ (\bar{3},3)$ Spinless Mesons}},\ }\href
  {https://doi.org/10.1103/PhysRevD.83.014015} {\bibfield  {journal} {\bibinfo
  {journal} {Phys. Rev. D}\ }\textbf {\bibinfo {volume} {83}},\ \bibinfo
  {pages} {014015} (\bibinfo {year} {2011})},\ \Eprint
  {https://arxiv.org/abs/1009.2422} {arXiv:1009.2422 [hep-ph]} \BibitemShut
  {NoStop}%
\bibitem [{\citenamefont {Chen}\ \emph {et~al.}(2012)\citenamefont {Chen},
  \citenamefont {Dmitrasinovic},\ and\ \citenamefont {Hosaka}}]{Chen:2011rh}%
  \BibitemOpen
  \bibfield  {author} {\bibinfo {author} {\bibfnamefont {H.-X.}\ \bibnamefont
  {Chen}}, \bibinfo {author} {\bibfnamefont {V.}~\bibnamefont
  {Dmitrasinovic}},\ and\ \bibinfo {author} {\bibfnamefont {A.}~\bibnamefont
  {Hosaka}},\ }\bibfield  {title} {\bibinfo {title} {{$mathrm{Baryons with}$
  $U_L(3) \times U_R(3)$ Chiral Symmetry IV: Interactions with Chiral (8,1)
  $\oplus$ (1,8) Vector and Axial-vector Mesons and Anomalous Magnetic
  Moments}},\ }\href {https://doi.org/10.1103/PhysRevC.85.055205} {\bibfield
  {journal} {\bibinfo  {journal} {Phys. Rev. C}\ }\textbf {\bibinfo {volume}
  {85}},\ \bibinfo {pages} {055205} (\bibinfo {year} {2012})},\ \Eprint
  {https://arxiv.org/abs/1109.3130} {arXiv:1109.3130 [hep-ph]} \BibitemShut
  {NoStop}%
\bibitem [{\citenamefont {Sasaki}\ \emph {et~al.}(2011)\citenamefont {Sasaki},
  \citenamefont {Lee}, \citenamefont {Paeng},\ and\ \citenamefont
  {Rho}}]{Sasaki:2011ff}%
  \BibitemOpen
  \bibfield  {author} {\bibinfo {author} {\bibfnamefont {C.}~\bibnamefont
  {Sasaki}}, \bibinfo {author} {\bibfnamefont {H.~K.}\ \bibnamefont {Lee}},
  \bibinfo {author} {\bibfnamefont {W.-G.}\ \bibnamefont {Paeng}},\ and\
  \bibinfo {author} {\bibfnamefont {M.}~\bibnamefont {Rho}},\ }\bibfield
  {title} {\bibinfo {title} {{Conformal anomaly and the vector coupling in
  dense matter}},\ }\href {https://doi.org/10.1103/PhysRevD.84.034011}
  {\bibfield  {journal} {\bibinfo  {journal} {Phys. Rev. D}\ }\textbf {\bibinfo
  {volume} {84}},\ \bibinfo {pages} {034011} (\bibinfo {year} {2011})},\
  \Eprint {https://arxiv.org/abs/1103.0184} {arXiv:1103.0184 [hep-ph]}
  \BibitemShut {NoStop}%
\bibitem [{\citenamefont {Motohiro}\ \emph {et~al.}(2015)\citenamefont
  {Motohiro}, \citenamefont {Kim},\ and\ \citenamefont
  {Harada}}]{Motohiro:2015taa}%
  \BibitemOpen
  \bibfield  {author} {\bibinfo {author} {\bibfnamefont {Y.}~\bibnamefont
  {Motohiro}}, \bibinfo {author} {\bibfnamefont {Y.}~\bibnamefont {Kim}},\ and\
  \bibinfo {author} {\bibfnamefont {M.}~\bibnamefont {Harada}},\ }\bibfield
  {title} {\bibinfo {title} {{Asymmetric nuclear matter in a parity doublet
  model with hidden local symmetry}},\ }\href
  {https://doi.org/10.1103/PhysRevC.92.025201} {\bibfield  {journal} {\bibinfo
  {journal} {Phys. Rev. C}\ }\textbf {\bibinfo {volume} {92}},\ \bibinfo
  {pages} {025201} (\bibinfo {year} {2015})},\ \bibinfo {note} {[Erratum:
  Phys.Rev.C 95, 059903 (2017)]},\ \Eprint {https://arxiv.org/abs/1505.00988}
  {arXiv:1505.00988 [nucl-th]} \BibitemShut {NoStop}%
\bibitem [{\citenamefont {Benic}\ \emph {et~al.}(2015)\citenamefont {Benic},
  \citenamefont {Mishustin},\ and\ \citenamefont {Sasaki}}]{Benic:2015pia}%
  \BibitemOpen
  \bibfield  {author} {\bibinfo {author} {\bibfnamefont {S.}~\bibnamefont
  {Benic}}, \bibinfo {author} {\bibfnamefont {I.}~\bibnamefont {Mishustin}},\
  and\ \bibinfo {author} {\bibfnamefont {C.}~\bibnamefont {Sasaki}},\
  }\bibfield  {title} {\bibinfo {title} {{Effective model for the QCD phase
  transitions at finite baryon density}},\ }\href
  {https://doi.org/10.1103/PhysRevD.91.125034} {\bibfield  {journal} {\bibinfo
  {journal} {Phys. Rev. D}\ }\textbf {\bibinfo {volume} {91}},\ \bibinfo
  {pages} {125034} (\bibinfo {year} {2015})},\ \Eprint
  {https://arxiv.org/abs/1502.05969} {arXiv:1502.05969 [hep-ph]} \BibitemShut
  {NoStop}%
\bibitem [{\citenamefont {Nishihara}\ and\ \citenamefont
  {Harada}(2015)}]{Nishihara:2015fka}%
  \BibitemOpen
  \bibfield  {author} {\bibinfo {author} {\bibfnamefont {H.}~\bibnamefont
  {Nishihara}}\ and\ \bibinfo {author} {\bibfnamefont {M.}~\bibnamefont
  {Harada}},\ }\bibfield  {title} {\bibinfo {title} {{Extended
  Goldberger-Treiman relation in a three-flavor parity doublet model}},\ }\href
  {https://doi.org/10.1103/PhysRevD.92.054022} {\bibfield  {journal} {\bibinfo
  {journal} {Phys. Rev. D}\ }\textbf {\bibinfo {volume} {92}},\ \bibinfo
  {pages} {054022} (\bibinfo {year} {2015})},\ \Eprint
  {https://arxiv.org/abs/1506.07956} {arXiv:1506.07956 [hep-ph]} \BibitemShut
  {NoStop}%
\bibitem [{\citenamefont {Suenaga}(2018)}]{Suenaga:2017wbb}%
  \BibitemOpen
  \bibfield  {author} {\bibinfo {author} {\bibfnamefont {D.}~\bibnamefont
  {Suenaga}},\ }\bibfield  {title} {\bibinfo {title} {{Examination of
  $N^*(1535)$ as a probe to observe the partial restoration of chiral symmetry
  in nuclear matter}},\ }\href {https://doi.org/10.1103/PhysRevC.97.045203}
  {\bibfield  {journal} {\bibinfo  {journal} {Phys. Rev. C}\ }\textbf {\bibinfo
  {volume} {97}},\ \bibinfo {pages} {045203} (\bibinfo {year} {2018})},\
  \Eprint {https://arxiv.org/abs/1704.03630} {arXiv:1704.03630 [nucl-th]}
  \BibitemShut {NoStop}%
\bibitem [{\citenamefont {Takeda}\ \emph {et~al.}(2018)\citenamefont {Takeda},
  \citenamefont {Kim},\ and\ \citenamefont {Harada}}]{Takeda:2017mrm}%
  \BibitemOpen
  \bibfield  {author} {\bibinfo {author} {\bibfnamefont {Y.}~\bibnamefont
  {Takeda}}, \bibinfo {author} {\bibfnamefont {Y.}~\bibnamefont {Kim}},\ and\
  \bibinfo {author} {\bibfnamefont {M.}~\bibnamefont {Harada}},\ }\bibfield
  {title} {\bibinfo {title} {{Catalysis of partial chiral symmetry restoration
  by $\Delta$ matter}},\ }\href {https://doi.org/10.1103/PhysRevC.97.065202}
  {\bibfield  {journal} {\bibinfo  {journal} {Phys. Rev. C}\ }\textbf {\bibinfo
  {volume} {97}},\ \bibinfo {pages} {065202} (\bibinfo {year} {2018})},\
  \Eprint {https://arxiv.org/abs/1704.04357} {arXiv:1704.04357 [nucl-th]}
  \BibitemShut {NoStop}%
\bibitem [{\citenamefont {Marczenko}\ and\ \citenamefont
  {Sasaki}(2018)}]{Marczenko:2017huu}%
  \BibitemOpen
  \bibfield  {author} {\bibinfo {author} {\bibfnamefont {M.}~\bibnamefont
  {Marczenko}}\ and\ \bibinfo {author} {\bibfnamefont {C.}~\bibnamefont
  {Sasaki}},\ }\bibfield  {title} {\bibinfo {title} {{Net-baryon number
  fluctuations in the Hybrid Quark-Meson-Nucleon model at finite density}},\
  }\href {https://doi.org/10.1103/PhysRevD.97.036011} {\bibfield  {journal}
  {\bibinfo  {journal} {Phys. Rev. D}\ }\textbf {\bibinfo {volume} {97}},\
  \bibinfo {pages} {036011} (\bibinfo {year} {2018})},\ \Eprint
  {https://arxiv.org/abs/1711.05521} {arXiv:1711.05521 [hep-ph]} \BibitemShut
  {NoStop}%
\bibitem [{\citenamefont {Mukherjee}\ \emph {et~al.}(2017)\citenamefont
  {Mukherjee}, \citenamefont {Schramm}, \citenamefont {Steinheimer},\ and\
  \citenamefont {Dexheimer}}]{Mukherjee:2017jzi}%
  \BibitemOpen
  \bibfield  {author} {\bibinfo {author} {\bibfnamefont {A.}~\bibnamefont
  {Mukherjee}}, \bibinfo {author} {\bibfnamefont {S.}~\bibnamefont {Schramm}},
  \bibinfo {author} {\bibfnamefont {J.}~\bibnamefont {Steinheimer}},\ and\
  \bibinfo {author} {\bibfnamefont {V.}~\bibnamefont {Dexheimer}},\ }\bibfield
  {title} {\bibinfo {title} {{The application of the Quark-Hadron Chiral
  Parity-Doublet Model to neutron star matter}},\ }\href
  {https://doi.org/10.1051/0004-6361/201731505} {\bibfield  {journal} {\bibinfo
   {journal} {Astron. Astrophys.}\ }\textbf {\bibinfo {volume} {608}},\
  \bibinfo {pages} {A110} (\bibinfo {year} {2017})},\ \Eprint
  {https://arxiv.org/abs/1706.09191} {arXiv:1706.09191 [nucl-th]} \BibitemShut
  {NoStop}%
\bibitem [{\citenamefont {Marczenko}\ \emph {et~al.}(2018)\citenamefont
  {Marczenko}, \citenamefont {Blaschke}, \citenamefont {Redlich},\ and\
  \citenamefont {Sasaki}}]{Marczenko:2018jui}%
  \BibitemOpen
  \bibfield  {author} {\bibinfo {author} {\bibfnamefont {M.}~\bibnamefont
  {Marczenko}}, \bibinfo {author} {\bibfnamefont {D.}~\bibnamefont {Blaschke}},
  \bibinfo {author} {\bibfnamefont {K.}~\bibnamefont {Redlich}},\ and\ \bibinfo
  {author} {\bibfnamefont {C.}~\bibnamefont {Sasaki}},\ }\bibfield  {title}
  {\bibinfo {title} {{Chiral symmetry restoration by parity doubling and the
  structure of neutron stars}},\ }\href
  {https://doi.org/10.1103/PhysRevD.98.103021} {\bibfield  {journal} {\bibinfo
  {journal} {Phys. Rev. D}\ }\textbf {\bibinfo {volume} {98}},\ \bibinfo
  {pages} {103021} (\bibinfo {year} {2018})},\ \Eprint
  {https://arxiv.org/abs/1805.06886} {arXiv:1805.06886 [nucl-th]} \BibitemShut
  {NoStop}%
\bibitem [{\citenamefont {Yamazaki}\ and\ \citenamefont
  {Harada}(2019{\natexlab{a}})}]{Yamazaki:2019tuo}%
  \BibitemOpen
  \bibfield  {author} {\bibinfo {author} {\bibfnamefont {T.}~\bibnamefont
  {Yamazaki}}\ and\ \bibinfo {author} {\bibfnamefont {M.}~\bibnamefont
  {Harada}},\ }\bibfield  {title} {\bibinfo {title} {{Constraint to chiral
  invariant masses of nucleons from GW170817 in an extended parity doublet
  model}},\ }\href {https://doi.org/10.1103/PhysRevC.100.025205} {\bibfield
  {journal} {\bibinfo  {journal} {Phys. Rev. C}\ }\textbf {\bibinfo {volume}
  {100}},\ \bibinfo {pages} {025205} (\bibinfo {year} {2019}{\natexlab{a}})},\
  \Eprint {https://arxiv.org/abs/1901.02167} {arXiv:1901.02167 [nucl-th]}
  \BibitemShut {NoStop}%
\bibitem [{\citenamefont {Harada}\ and\ \citenamefont
  {Yamazaki}(2019)}]{Harada:2019oaq}%
  \BibitemOpen
  \bibfield  {author} {\bibinfo {author} {\bibfnamefont {M.}~\bibnamefont
  {Harada}}\ and\ \bibinfo {author} {\bibfnamefont {T.}~\bibnamefont
  {Yamazaki}},\ }\bibfield  {title} {\bibinfo {title} {{Charmed Mesons in
  Nuclear Matter Based on Chiral Effective Models}},\ }\href
  {https://doi.org/10.7566/JPSCP.26.024001} {\bibfield  {journal} {\bibinfo
  {journal} {JPS Conf. Proc.}\ }\textbf {\bibinfo {volume} {26}},\ \bibinfo
  {pages} {024001} (\bibinfo {year} {2019})}\BibitemShut {NoStop}%
\bibitem [{\citenamefont {Harada}(2020)}]{Harada:2020etl}%
  \BibitemOpen
  \bibfield  {author} {\bibinfo {author} {\bibfnamefont {M.}~\bibnamefont
  {Harada}},\ }\bibfield  {title} {\bibinfo {title} {{Dense nuclear matter
  based on a chiral model with parity doublet structure}},\ }in\ \href
  {https://doi.org/10.1142/9789811219313_0113} {\emph {\bibinfo {booktitle}
  {{18th International Conference on Hadron Spectroscopy and Structure}}}}\
  (\bibinfo {year} {2020})\ pp.\ \bibinfo {pages} {661--666}\BibitemShut
  {NoStop}%
\bibitem [{\citenamefont {Marczenko}(2020)}]{Marczenko:2020yok}%
  \BibitemOpen
  \bibfield  {author} {\bibinfo {author} {\bibfnamefont {M.}~\bibnamefont
  {Marczenko}},\ }\bibfield  {title} {\bibinfo {title} {{Hybrid quark-hadron
  equation of state for multi-messenger astronomy}},\ }in\ \href@noop {} {\emph
  {\bibinfo {booktitle} {{Criticality in QCD and the Hadron Resonance Gas}}}}\
  (\bibinfo {year} {2020})\ \Eprint {https://arxiv.org/abs/2010.15420}
  {arXiv:2010.15420 [nucl-th]} \BibitemShut {NoStop}%
\bibitem [{\citenamefont {Marczenko}\ \emph {et~al.}(2021)\citenamefont
  {Marczenko}, \citenamefont {Redlich},\ and\ \citenamefont
  {Sasaki}}]{Marczenko:2021icv}%
  \BibitemOpen
  \bibfield  {author} {\bibinfo {author} {\bibfnamefont {M.}~\bibnamefont
  {Marczenko}}, \bibinfo {author} {\bibfnamefont {K.}~\bibnamefont {Redlich}},\
  and\ \bibinfo {author} {\bibfnamefont {C.}~\bibnamefont {Sasaki}},\
  }\bibfield  {title} {\bibinfo {title} {{Interplay between chiral dynamics and
  repulsive interactions in hot hadronic matter}},\ }\href
  {https://doi.org/10.1103/physrevd.103.054035} {\bibfield  {journal} {\bibinfo
   {journal} {Phys. Rev. D}\ }\textbf {\bibinfo {volume} {103}},\ \bibinfo
  {pages} {054035} (\bibinfo {year} {2021})}\BibitemShut {NoStop}%
\bibitem [{\citenamefont {Minamikawa}\ \emph
  {et~al.}(2023{\natexlab{a}})\citenamefont {Minamikawa}, \citenamefont {Gao},
  \citenamefont {kojo},\ and\ \citenamefont {Harada}}]{Minamikawa:2023ypn}%
  \BibitemOpen
  \bibfield  {author} {\bibinfo {author} {\bibfnamefont {T.}~\bibnamefont
  {Minamikawa}}, \bibinfo {author} {\bibfnamefont {B.}~\bibnamefont {Gao}},
  \bibinfo {author} {\bibfnamefont {T.}~\bibnamefont {kojo}},\ and\ \bibinfo
  {author} {\bibfnamefont {M.}~\bibnamefont {Harada}},\ }\bibfield  {title}
  {\bibinfo {title} {{Parity doublet model for baryon octets: Diquark
  classifications and mass hierarchy based on the quark-line diagram}},\ }\href
  {https://doi.org/10.1103/PhysRevD.108.076017} {\bibfield  {journal} {\bibinfo
   {journal} {Phys. Rev. D}\ }\textbf {\bibinfo {volume} {108}},\ \bibinfo
  {pages} {076017} (\bibinfo {year} {2023}{\natexlab{a}})},\ \Eprint
  {https://arxiv.org/abs/2306.15564} {arXiv:2306.15564 [hep-ph]} \BibitemShut
  {NoStop}%
\bibitem [{\citenamefont {Gao}\ \emph {et~al.}(2024{\natexlab{a}})\citenamefont
  {Gao}, \citenamefont {Kojo},\ and\ \citenamefont {Harada}}]{Gao:2024mew}%
  \BibitemOpen
  \bibfield  {author} {\bibinfo {author} {\bibfnamefont {B.}~\bibnamefont
  {Gao}}, \bibinfo {author} {\bibfnamefont {T.}~\bibnamefont {Kojo}},\ and\
  \bibinfo {author} {\bibfnamefont {M.}~\bibnamefont {Harada}},\ }\bibfield
  {title} {\bibinfo {title} {{Parity doublet model for baryon octets: Ground
  states saturated by good diquarks and the role of bad diquarks for excited
  states}},\ }\href {https://doi.org/10.1103/PhysRevD.110.016016} {\bibfield
  {journal} {\bibinfo  {journal} {Phys. Rev. D}\ }\textbf {\bibinfo {volume}
  {110}},\ \bibinfo {pages} {016016} (\bibinfo {year} {2024}{\natexlab{a}})},\
  \Eprint {https://arxiv.org/abs/2403.18214} {arXiv:2403.18214 [hep-ph]}
  \BibitemShut {NoStop}%
\bibitem [{\citenamefont {Christos}(1983)}]{Christos:1982kc}%
  \BibitemOpen
  \bibfield  {author} {\bibinfo {author} {\bibfnamefont {G.~A.}\ \bibnamefont
  {Christos}},\ }\bibfield  {title} {\bibinfo {title} {{Effective Chiral
  Lagrangians With Baryons: The Mass Splitting of the Spin 1/2 Parity
  Partners}},\ }\href {https://doi.org/10.1007/BF01648780} {\bibfield
  {journal} {\bibinfo  {journal} {Z. Phys. C}\ }\textbf {\bibinfo {volume}
  {21}},\ \bibinfo {pages} {83} (\bibinfo {year} {1983})}\BibitemShut {NoStop}%
\bibitem [{\citenamefont {Jaffe}(1977{\natexlab{a}})}]{Jaffe:1976ig}%
  \BibitemOpen
  \bibfield  {author} {\bibinfo {author} {\bibfnamefont {R.~L.}\ \bibnamefont
  {Jaffe}},\ }\bibfield  {title} {\bibinfo {title} {{Multi-Quark Hadrons. 1.
  The Phenomenology of (2 Quark 2 anti-Quark) Mesons}},\ }\href
  {https://doi.org/10.1103/PhysRevD.15.267} {\bibfield  {journal} {\bibinfo
  {journal} {Phys. Rev. D}\ }\textbf {\bibinfo {volume} {15}},\ \bibinfo
  {pages} {267} (\bibinfo {year} {1977}{\natexlab{a}})}\BibitemShut {NoStop}%
\bibitem [{\citenamefont {Jaffe}(1977{\natexlab{b}})}]{Jaffe:1976ih}%
  \BibitemOpen
  \bibfield  {author} {\bibinfo {author} {\bibfnamefont {R.~L.}\ \bibnamefont
  {Jaffe}},\ }\bibfield  {title} {\bibinfo {title} {{Multi-Quark Hadrons. 2.
  Methods}},\ }\href {https://doi.org/10.1103/PhysRevD.15.281} {\bibfield
  {journal} {\bibinfo  {journal} {Phys. Rev. D}\ }\textbf {\bibinfo {volume}
  {15}},\ \bibinfo {pages} {281} (\bibinfo {year}
  {1977}{\natexlab{b}})}\BibitemShut {NoStop}%
\bibitem [{\citenamefont {Rapp}\ \emph {et~al.}(1998)\citenamefont {Rapp},
  \citenamefont {Sch\"afer}, \citenamefont {Shuryak},\ and\ \citenamefont
  {Velkovsky}}]{Rapp:1997zu}%
  \BibitemOpen
  \bibfield  {author} {\bibinfo {author} {\bibfnamefont {R.}~\bibnamefont
  {Rapp}}, \bibinfo {author} {\bibfnamefont {T.}~\bibnamefont {Sch\"afer}},
  \bibinfo {author} {\bibfnamefont {E.~V.}\ \bibnamefont {Shuryak}},\ and\
  \bibinfo {author} {\bibfnamefont {M.}~\bibnamefont {Velkovsky}},\ }\bibfield
  {title} {\bibinfo {title} {{Diquark Bose condensates in high density matter
  and instantons}},\ }\href {https://doi.org/10.1103/PhysRevLett.81.53}
  {\bibfield  {journal} {\bibinfo  {journal} {Phys. Rev. Lett.}\ }\textbf
  {\bibinfo {volume} {81}},\ \bibinfo {pages} {53} (\bibinfo {year} {1998})},\
  \Eprint {https://arxiv.org/abs/hep-ph/9711396} {arXiv:hep-ph/9711396}
  \BibitemShut {NoStop}%
\bibitem [{\citenamefont {Jaffe}\ and\ \citenamefont
  {Wilczek}(2003)}]{Jaffe:2003sg}%
  \BibitemOpen
  \bibfield  {author} {\bibinfo {author} {\bibfnamefont {R.~L.}\ \bibnamefont
  {Jaffe}}\ and\ \bibinfo {author} {\bibfnamefont {F.}~\bibnamefont
  {Wilczek}},\ }\bibfield  {title} {\bibinfo {title} {{Diquarks and exotic
  spectroscopy}},\ }\href {https://doi.org/10.1103/PhysRevLett.91.232003}
  {\bibfield  {journal} {\bibinfo  {journal} {Phys. Rev. Lett.}\ }\textbf
  {\bibinfo {volume} {91}},\ \bibinfo {pages} {232003} (\bibinfo {year}
  {2003})},\ \Eprint {https://arxiv.org/abs/hep-ph/0307341}
  {arXiv:hep-ph/0307341} \BibitemShut {NoStop}%
\bibitem [{\citenamefont {Eichmann}\ \emph {et~al.}(2016)\citenamefont
  {Eichmann}, \citenamefont {Sanchis-Alepuz}, \citenamefont {Williams},
  \citenamefont {Alkofer},\ and\ \citenamefont {Fischer}}]{Eichmann:2016yit}%
  \BibitemOpen
  \bibfield  {author} {\bibinfo {author} {\bibfnamefont {G.}~\bibnamefont
  {Eichmann}}, \bibinfo {author} {\bibfnamefont {H.}~\bibnamefont
  {Sanchis-Alepuz}}, \bibinfo {author} {\bibfnamefont {R.}~\bibnamefont
  {Williams}}, \bibinfo {author} {\bibfnamefont {R.}~\bibnamefont {Alkofer}},\
  and\ \bibinfo {author} {\bibfnamefont {C.~S.}\ \bibnamefont {Fischer}},\
  }\bibfield  {title} {\bibinfo {title} {{Baryons as relativistic three-quark
  bound states}},\ }\href {https://doi.org/10.1016/j.ppnp.2016.07.001}
  {\bibfield  {journal} {\bibinfo  {journal} {Prog. Part. Nucl. Phys.}\
  }\textbf {\bibinfo {volume} {91}},\ \bibinfo {pages} {1} (\bibinfo {year}
  {2016})},\ \Eprint {https://arxiv.org/abs/1606.09602} {arXiv:1606.09602
  [hep-ph]} \BibitemShut {NoStop}%
\bibitem [{\citenamefont {Chen}\ \emph {et~al.}(2023)\citenamefont {Chen},
  \citenamefont {Chen}, \citenamefont {Liu}, \citenamefont {Liu},\ and\
  \citenamefont {Zhu}}]{Chen:2022asf}%
  \BibitemOpen
  \bibfield  {author} {\bibinfo {author} {\bibfnamefont {H.-X.}\ \bibnamefont
  {Chen}}, \bibinfo {author} {\bibfnamefont {W.}~\bibnamefont {Chen}}, \bibinfo
  {author} {\bibfnamefont {X.}~\bibnamefont {Liu}}, \bibinfo {author}
  {\bibfnamefont {Y.-R.}\ \bibnamefont {Liu}},\ and\ \bibinfo {author}
  {\bibfnamefont {S.-L.}\ \bibnamefont {Zhu}},\ }\bibfield  {title} {\bibinfo
  {title} {{An updated review of the new hadron states}},\ }\href
  {https://doi.org/10.1088/1361-6633/aca3b6} {\bibfield  {journal} {\bibinfo
  {journal} {Rept. Prog. Phys.}\ }\textbf {\bibinfo {volume} {86}},\ \bibinfo
  {pages} {026201} (\bibinfo {year} {2023})},\ \Eprint
  {https://arxiv.org/abs/2204.02649} {arXiv:2204.02649 [hep-ph]} \BibitemShut
  {NoStop}%
\bibitem [{\citenamefont {Navas}\ \emph {et~al.}(2024)\citenamefont {Navas}
  \emph {et~al.}}]{ParticleDataGroup:2024cfk}%
  \BibitemOpen
  \bibfield  {author} {\bibinfo {author} {\bibfnamefont {S.}~\bibnamefont
  {Navas}} \emph {et~al.} (\bibinfo {collaboration} {Particle Data Group}),\
  }\bibfield  {title} {\bibinfo {title} {{Review of particle physics}},\ }\href
  {https://doi.org/10.1103/PhysRevD.110.030001} {\bibfield  {journal} {\bibinfo
   {journal} {Phys. Rev. D}\ }\textbf {\bibinfo {volume} {110}},\ \bibinfo
  {pages} {030001} (\bibinfo {year} {2024})}\BibitemShut {NoStop}%
\bibitem [{\citenamefont {Kaiser}\ \emph {et~al.}(1995)\citenamefont {Kaiser},
  \citenamefont {Siegel},\ and\ \citenamefont {Weise}}]{Kaiser:1995cy}%
  \BibitemOpen
  \bibfield  {author} {\bibinfo {author} {\bibfnamefont {N.}~\bibnamefont
  {Kaiser}}, \bibinfo {author} {\bibfnamefont {P.~B.}\ \bibnamefont {Siegel}},\
  and\ \bibinfo {author} {\bibfnamefont {W.}~\bibnamefont {Weise}},\ }\bibfield
   {title} {\bibinfo {title} {{Chiral dynamics and the S11 (1535) nucleon
  resonance}},\ }\href {https://doi.org/10.1016/0370-2693(95)01203-3}
  {\bibfield  {journal} {\bibinfo  {journal} {Phys. Lett. B}\ }\textbf
  {\bibinfo {volume} {362}},\ \bibinfo {pages} {23} (\bibinfo {year} {1995})},\
  \Eprint {https://arxiv.org/abs/nucl-th/9507036} {arXiv:nucl-th/9507036}
  \BibitemShut {NoStop}%
\bibitem [{\citenamefont {Kaiser}\ \emph {et~al.}(1997)\citenamefont {Kaiser},
  \citenamefont {Waas},\ and\ \citenamefont {Weise}}]{Kaiser:1996js}%
  \BibitemOpen
  \bibfield  {author} {\bibinfo {author} {\bibfnamefont {N.}~\bibnamefont
  {Kaiser}}, \bibinfo {author} {\bibfnamefont {T.}~\bibnamefont {Waas}},\ and\
  \bibinfo {author} {\bibfnamefont {W.}~\bibnamefont {Weise}},\ }\bibfield
  {title} {\bibinfo {title} {{SU(3) chiral dynamics with coupled channels: Eta
  and kaon photoproduction}},\ }\href
  {https://doi.org/10.1016/S0375-9474(96)00321-1} {\bibfield  {journal}
  {\bibinfo  {journal} {Nucl. Phys. A}\ }\textbf {\bibinfo {volume} {612}},\
  \bibinfo {pages} {297} (\bibinfo {year} {1997})},\ \Eprint
  {https://arxiv.org/abs/hep-ph/9607459} {arXiv:hep-ph/9607459} \BibitemShut
  {NoStop}%
\bibitem [{\citenamefont {Inoue}\ \emph {et~al.}(2002)\citenamefont {Inoue},
  \citenamefont {Oset},\ and\ \citenamefont {Vicente~Vacas}}]{Inoue:2001ip}%
  \BibitemOpen
  \bibfield  {author} {\bibinfo {author} {\bibfnamefont {T.}~\bibnamefont
  {Inoue}}, \bibinfo {author} {\bibfnamefont {E.}~\bibnamefont {Oset}},\ and\
  \bibinfo {author} {\bibfnamefont {M.~J.}\ \bibnamefont {Vicente~Vacas}},\
  }\bibfield  {title} {\bibinfo {title} {{Chiral unitary approach to S wave
  meson baryon scattering in the strangeness S = O sector}},\ }\href
  {https://doi.org/10.1103/PhysRevC.65.035204} {\bibfield  {journal} {\bibinfo
  {journal} {Phys. Rev. C}\ }\textbf {\bibinfo {volume} {65}},\ \bibinfo
  {pages} {035204} (\bibinfo {year} {2002})},\ \Eprint
  {https://arxiv.org/abs/hep-ph/0110333} {arXiv:hep-ph/0110333} \BibitemShut
  {NoStop}%
\bibitem [{\citenamefont {Ramos}\ \emph {et~al.}(2002)\citenamefont {Ramos},
  \citenamefont {Oset},\ and\ \citenamefont {Bennhold}}]{Ramos:2002xh}%
  \BibitemOpen
  \bibfield  {author} {\bibinfo {author} {\bibfnamefont {A.}~\bibnamefont
  {Ramos}}, \bibinfo {author} {\bibfnamefont {E.}~\bibnamefont {Oset}},\ and\
  \bibinfo {author} {\bibfnamefont {C.}~\bibnamefont {Bennhold}},\ }\bibfield
  {title} {\bibinfo {title} {{On the spin, parity and nature of the Xi(1620)
  resonance}},\ }\href {https://doi.org/10.1103/PhysRevLett.89.252001}
  {\bibfield  {journal} {\bibinfo  {journal} {Phys. Rev. Lett.}\ }\textbf
  {\bibinfo {volume} {89}},\ \bibinfo {pages} {252001} (\bibinfo {year}
  {2002})},\ \Eprint {https://arxiv.org/abs/nucl-th/0204044}
  {arXiv:nucl-th/0204044} \BibitemShut {NoStop}%
\bibitem [{\citenamefont {Oh}(2007)}]{Oh:2007cr}%
  \BibitemOpen
  \bibfield  {author} {\bibinfo {author} {\bibfnamefont {Y.}~\bibnamefont
  {Oh}},\ }\bibfield  {title} {\bibinfo {title} {{Xi and Omega baryons in the
  Skyrme model}},\ }\href {https://doi.org/10.1103/PhysRevD.75.074002}
  {\bibfield  {journal} {\bibinfo  {journal} {Phys. Rev. D}\ }\textbf {\bibinfo
  {volume} {75}},\ \bibinfo {pages} {074002} (\bibinfo {year} {2007})},\
  \Eprint {https://arxiv.org/abs/hep-ph/0702126} {arXiv:hep-ph/0702126}
  \BibitemShut {NoStop}%
\bibitem [{\citenamefont {Huang}\ and\ \citenamefont
  {Geng}(2020)}]{Huang:2020taj}%
  \BibitemOpen
  \bibfield  {author} {\bibinfo {author} {\bibfnamefont {Y.}~\bibnamefont
  {Huang}}\ and\ \bibinfo {author} {\bibfnamefont {L.}~\bibnamefont {Geng}},\
  }\bibfield  {title} {\bibinfo {title} {{Strong decays of the $\Xi(1620)$ as a
  $\Lambda\bar{K}$ and $\Sigma\bar{K}$ molecule}},\ }\href
  {https://doi.org/10.1140/epjc/s10052-020-8421-9} {\bibfield  {journal}
  {\bibinfo  {journal} {Eur. Phys. J. C}\ }\textbf {\bibinfo {volume} {80}},\
  \bibinfo {pages} {837} (\bibinfo {year} {2020})},\ \Eprint
  {https://arxiv.org/abs/2006.06227} {arXiv:2006.06227 [hep-ph]} \BibitemShut
  {NoStop}%
\bibitem [{\citenamefont {Sekihara}(2015)}]{Sekihara:2015qqa}%
  \BibitemOpen
  \bibfield  {author} {\bibinfo {author} {\bibfnamefont {T.}~\bibnamefont
  {Sekihara}},\ }\bibfield  {title} {\bibinfo {title} {{$\Xi (1690)$ as a
  $\bar{K} \Sigma$ molecular state}},\ }\href
  {https://doi.org/10.1093/ptep/ptv129} {\bibfield  {journal} {\bibinfo
  {journal} {PTEP}\ }\textbf {\bibinfo {volume} {2015}},\ \bibinfo {pages}
  {091D01} (\bibinfo {year} {2015})},\ \Eprint
  {https://arxiv.org/abs/1505.02849} {arXiv:1505.02849 [hep-ph]} \BibitemShut
  {NoStop}%
\bibitem [{\citenamefont {Sekihara}(2017)}]{Sekihara:2016hcf}%
  \BibitemOpen
  \bibfield  {author} {\bibinfo {author} {\bibfnamefont {T.}~\bibnamefont
  {Sekihara}},\ }\bibfield  {title} {\bibinfo {title} {{Dynamically Generated
  $\Xi(1690)$}},\ }\href {https://doi.org/10.7566/JPSCP.17.072007} {\bibfield
  {journal} {\bibinfo  {journal} {JPS Conf. Proc.}\ }\textbf {\bibinfo {volume}
  {17}},\ \bibinfo {pages} {072007} (\bibinfo {year} {2017})},\ \Eprint
  {https://arxiv.org/abs/1607.03374} {arXiv:1607.03374 [hep-ph]} \BibitemShut
  {NoStop}%
\bibitem [{\citenamefont {Xiao}\ and\ \citenamefont
  {Zhong}(2013)}]{Xiao:2013xi}%
  \BibitemOpen
  \bibfield  {author} {\bibinfo {author} {\bibfnamefont {L.-Y.}\ \bibnamefont
  {Xiao}}\ and\ \bibinfo {author} {\bibfnamefont {X.-H.}\ \bibnamefont
  {Zhong}},\ }\bibfield  {title} {\bibinfo {title} {{$\Xi$ baryon strong decays
  in a chiral quark model}},\ }\href
  {https://doi.org/10.1103/PhysRevD.87.094002} {\bibfield  {journal} {\bibinfo
  {journal} {Phys. Rev. D}\ }\textbf {\bibinfo {volume} {87}},\ \bibinfo
  {pages} {094002} (\bibinfo {year} {2013})},\ \Eprint
  {https://arxiv.org/abs/1302.0079} {arXiv:1302.0079 [hep-ph]} \BibitemShut
  {NoStop}%
\bibitem [{\citenamefont {Minamikawa}\ \emph {et~al.}(2021)\citenamefont
  {Minamikawa}, \citenamefont {Kojo},\ and\ \citenamefont
  {Harada}}]{Minamikawa:2020jfj}%
  \BibitemOpen
  \bibfield  {author} {\bibinfo {author} {\bibfnamefont {T.}~\bibnamefont
  {Minamikawa}}, \bibinfo {author} {\bibfnamefont {T.}~\bibnamefont {Kojo}},\
  and\ \bibinfo {author} {\bibfnamefont {M.}~\bibnamefont {Harada}},\
  }\bibfield  {title} {\bibinfo {title} {{Quark-hadron crossover equations of
  state for neutron stars: constraining the chiral invariant mass in a parity
  doublet model}},\ }\href {https://doi.org/10.1103/PhysRevC.103.045205}
  {\bibfield  {journal} {\bibinfo  {journal} {Phys. Rev. C}\ }\textbf {\bibinfo
  {volume} {103}},\ \bibinfo {pages} {045205} (\bibinfo {year} {2021})},\
  \Eprint {https://arxiv.org/abs/2011.13684} {arXiv:2011.13684 [nucl-th]}
  \BibitemShut {NoStop}%
\bibitem [{\citenamefont {Minamikawa}\ \emph
  {et~al.}(2023{\natexlab{b}})\citenamefont {Minamikawa}, \citenamefont {Gao},
  \citenamefont {Kojo},\ and\ \citenamefont {Harada}}]{Minamikawa:2023eky}%
  \BibitemOpen
  \bibfield  {author} {\bibinfo {author} {\bibfnamefont {T.}~\bibnamefont
  {Minamikawa}}, \bibinfo {author} {\bibfnamefont {B.}~\bibnamefont {Gao}},
  \bibinfo {author} {\bibfnamefont {T.}~\bibnamefont {Kojo}},\ and\ \bibinfo
  {author} {\bibfnamefont {M.}~\bibnamefont {Harada}},\ }\bibfield  {title}
  {\bibinfo {title} {{Chiral Restoration of Nucleons in Neutron Star Matter:
  Studies Based on a Parity Doublet Model}},\ }\href
  {https://doi.org/10.3390/sym15030745} {\bibfield  {journal} {\bibinfo
  {journal} {Symmetry}\ }\textbf {\bibinfo {volume} {15}},\ \bibinfo {pages}
  {745} (\bibinfo {year} {2023}{\natexlab{b}})},\ \Eprint
  {https://arxiv.org/abs/2302.00825} {arXiv:2302.00825 [nucl-th]} \BibitemShut
  {NoStop}%
\bibitem [{\citenamefont {Kong}\ \emph {et~al.}(2023)\citenamefont {Kong},
  \citenamefont {Minamikawa},\ and\ \citenamefont {Harada}}]{Kong:2023nue}%
  \BibitemOpen
  \bibfield  {author} {\bibinfo {author} {\bibfnamefont {Y.~K.}\ \bibnamefont
  {Kong}}, \bibinfo {author} {\bibfnamefont {T.}~\bibnamefont {Minamikawa}},\
  and\ \bibinfo {author} {\bibfnamefont {M.}~\bibnamefont {Harada}},\
  }\bibfield  {title} {\bibinfo {title} {{Neutron star matter based on a parity
  doublet model including the a0(980) meson}},\ }\href
  {https://doi.org/10.1103/PhysRevC.108.055206} {\bibfield  {journal} {\bibinfo
   {journal} {Phys. Rev. C}\ }\textbf {\bibinfo {volume} {108}},\ \bibinfo
  {pages} {055206} (\bibinfo {year} {2023})},\ \Eprint
  {https://arxiv.org/abs/2306.08140} {arXiv:2306.08140 [nucl-th]} \BibitemShut
  {NoStop}%
\bibitem [{\citenamefont {Gao}\ \emph {et~al.}(2024{\natexlab{b}})\citenamefont
  {Gao}, \citenamefont {Yan},\ and\ \citenamefont {Harada}}]{Gao:2024chh}%
  \BibitemOpen
  \bibfield  {author} {\bibinfo {author} {\bibfnamefont {B.}~\bibnamefont
  {Gao}}, \bibinfo {author} {\bibfnamefont {Y.}~\bibnamefont {Yan}},\ and\
  \bibinfo {author} {\bibfnamefont {M.}~\bibnamefont {Harada}},\ }\bibfield
  {title} {\bibinfo {title} {{Reconciling constraints from the supernova
  remnant HESS J1731-347 with the parity doublet model}},\ }\href
  {https://doi.org/10.1103/PhysRevC.109.065807} {\bibfield  {journal} {\bibinfo
   {journal} {Phys. Rev. C}\ }\textbf {\bibinfo {volume} {109}},\ \bibinfo
  {pages} {065807} (\bibinfo {year} {2024}{\natexlab{b}})},\ \Eprint
  {https://arxiv.org/abs/2404.04786} {arXiv:2404.04786 [nucl-th]} \BibitemShut
  {NoStop}%
\bibitem [{\citenamefont {Kong}\ \emph {et~al.}(2025)\citenamefont {Kong},
  \citenamefont {Gao},\ and\ \citenamefont {Harada}}]{Kong:2025dwl}%
  \BibitemOpen
  \bibfield  {author} {\bibinfo {author} {\bibfnamefont {Y.-K.}\ \bibnamefont
  {Kong}}, \bibinfo {author} {\bibfnamefont {B.}~\bibnamefont {Gao}},\ and\
  \bibinfo {author} {\bibfnamefont {M.}~\bibnamefont {Harada}},\ }\bibfield
  {title} {\bibinfo {title} {{Chiral Invariant Mass Constraints from HESS J1731
  347 in an Extended Parity Doublet Model with Isovector Scalar Meson}},\
  }\href@noop {} {\  (\bibinfo {year} {2025})},\ \Eprint
  {https://arxiv.org/abs/2506.16684} {arXiv:2506.16684 [nucl-th]} \BibitemShut
  {NoStop}%
\bibitem [{\citenamefont {Gao}\ \emph {et~al.}(2025{\natexlab{a}})\citenamefont
  {Gao}, \citenamefont {Liu}, \citenamefont {Harada},\ and\ \citenamefont
  {Ma}}]{Gao:2025nkg}%
  \BibitemOpen
  \bibfield  {author} {\bibinfo {author} {\bibfnamefont {B.}~\bibnamefont
  {Gao}}, \bibinfo {author} {\bibfnamefont {X.}~\bibnamefont {Liu}}, \bibinfo
  {author} {\bibfnamefont {M.}~\bibnamefont {Harada}},\ and\ \bibinfo {author}
  {\bibfnamefont {Y.-L.}\ \bibnamefont {Ma}},\ }\bibfield  {title} {\bibinfo
  {title} {{Implication of neutron star observations to the origin of nucleon
  mass}},\ }\href@noop {} {\  (\bibinfo {year} {2025}{\natexlab{a}})},\ \Eprint
  {https://arxiv.org/abs/2508.00243} {arXiv:2508.00243 [nucl-th]} \BibitemShut
  {NoStop}%
\bibitem [{\citenamefont {Gao}\ \emph {et~al.}(2025{\natexlab{b}})\citenamefont
  {Gao}, \citenamefont {Kong},\ and\ \citenamefont {Ma}}]{Gao:2025vdc}%
  \BibitemOpen
  \bibfield  {author} {\bibinfo {author} {\bibfnamefont {B.}~\bibnamefont
  {Gao}}, \bibinfo {author} {\bibfnamefont {Y.-K.}\ \bibnamefont {Kong}},\ and\
  \bibinfo {author} {\bibfnamefont {Y.-L.}\ \bibnamefont {Ma}},\ }\bibfield
  {title} {\bibinfo {title} {{Origin of nucleon mass in the light of PSR
  J0614-3329 with quark-hadron crossover}},\ }\href
  {https://doi.org/10.1103/xwkv-s8lg} {\bibfield  {journal} {\bibinfo
  {journal} {Phys. Rev. D}\ }\textbf {\bibinfo {volume} {112}},\ \bibinfo
  {pages} {083041} (\bibinfo {year} {2025}{\natexlab{b}})},\ \Eprint
  {https://arxiv.org/abs/2509.03008} {arXiv:2509.03008 [nucl-th]} \BibitemShut
  {NoStop}%
\bibitem [{\citenamefont {Jido}\ \emph
  {et~al.}(2000{\natexlab{b}})\citenamefont {Jido}, \citenamefont {Hatsuda},\
  and\ \citenamefont {Kunihiro}}]{Jido:1999hd}%
  \BibitemOpen
  \bibfield  {author} {\bibinfo {author} {\bibfnamefont {D.}~\bibnamefont
  {Jido}}, \bibinfo {author} {\bibfnamefont {T.}~\bibnamefont {Hatsuda}},\ and\
  \bibinfo {author} {\bibfnamefont {T.}~\bibnamefont {Kunihiro}},\ }\bibfield
  {title} {\bibinfo {title} {{Chiral symmetry realization for even parity and
  odd parity baryon resonances}},\ }\href
  {https://doi.org/10.1103/PhysRevLett.84.3252} {\bibfield  {journal} {\bibinfo
   {journal} {Phys. Rev. Lett.}\ }\textbf {\bibinfo {volume} {84}},\ \bibinfo
  {pages} {3252} (\bibinfo {year} {2000}{\natexlab{b}})},\ \Eprint
  {https://arxiv.org/abs/hep-ph/9910375} {arXiv:hep-ph/9910375} \BibitemShut
  {NoStop}%
\bibitem [{\citenamefont {Yamazaki}\ and\ \citenamefont
  {Harada}(2019{\natexlab{b}})}]{Yamazaki:2018stk}%
  \BibitemOpen
  \bibfield  {author} {\bibinfo {author} {\bibfnamefont {T.}~\bibnamefont
  {Yamazaki}}\ and\ \bibinfo {author} {\bibfnamefont {M.}~\bibnamefont
  {Harada}},\ }\bibfield  {title} {\bibinfo {title} {{Chiral partner structure
  of light nucleons in an extended parity doublet model}},\ }\href
  {https://doi.org/10.1103/PhysRevD.99.034012} {\bibfield  {journal} {\bibinfo
  {journal} {Phys. Rev. D}\ }\textbf {\bibinfo {volume} {99}},\ \bibinfo
  {pages} {034012} (\bibinfo {year} {2019}{\natexlab{b}})},\ \Eprint
  {https://arxiv.org/abs/1809.02359} {arXiv:1809.02359 [hep-ph]} \BibitemShut
  {NoStop}%
\bibitem [{\citenamefont {Kummer}\ \emph {et~al.}(2025)\citenamefont {Kummer},
  \citenamefont {Leupold},\ and\ \citenamefont {von Smekal}}]{Kummer:2025kch}%
  \BibitemOpen
  \bibfield  {author} {\bibinfo {author} {\bibfnamefont {C.}~\bibnamefont
  {Kummer}}, \bibinfo {author} {\bibfnamefont {S.}~\bibnamefont {Leupold}},\
  and\ \bibinfo {author} {\bibfnamefont {L.}~\bibnamefont {von Smekal}},\
  }\bibfield  {title} {\bibinfo {title} {{Kinetic Mixing and Axial Charges in
  the Parity-Doublet Model}},\ }\href@noop {} {\  (\bibinfo {year} {2025})},\
  \Eprint {https://arxiv.org/abs/2512.03894} {arXiv:2512.03894 [hep-ph]}
  \BibitemShut {NoStop}%
\end{thebibliography}%

\end{document}